\documentclass[sigconf]{acmart}
\usepackage{amsmath,amsfonts}
\usepackage{bm}            
\usepackage{mathtools}     
\usepackage[linesnumbered,ruled,vlined]{algorithm2e} 

\usepackage{booktabs}    
\usepackage{multirow}    
\usepackage{colortbl}    
\usepackage[table]{xcolor} 
\usepackage{hyperref}
\usepackage{tabularx}

\usepackage{pifont}

\usepackage[most]{tcolorbox}
\usepackage{xcolor}
\usepackage{tikz}
\usetikzlibrary{calc}
\usepackage{mathtools}
\tcbuselibrary{skins, breakable}

\definecolor{twDark}{HTML}{0A4A4A}   
\definecolor{twBorder}{HTML}{0A4A4A} 
\definecolor{twFill}{HTML}{F4FBFA}   
\definecolor{tbGrey}{HTML}{F3F3F3}
\definecolor{tbRed}{HTML}{FCE8E8}

\newcolumntype{Y}{>{\centering\arraybackslash}X}
\newcommand{\cnum}[1]{\ding{\numexpr181+#1}}
\newtcolorbox{takeawaybox}[1][]{%
  enhanced,
  breakable,
  colback=twFill,
  colframe=twBorder,
  boxrule=0.6pt,
  arc=8pt,
  left=8pt,
  right=8pt,
  top=6pt,
  bottom=2pt,
  before skip=10pt,
  after skip=10pt,
    overlay={%
      \node[
        fill=twDark,
        text=white,
        font=\sffamily\bfseries\footnotesize,
        rounded corners=6pt,
        inner xsep=8pt,
        inner ysep=3pt,
        anchor=north west 
      ] at ([xshift=16pt,yshift=4pt]frame.north west) {#1}; 
    },
}

\usepackage{graphicx}
\usepackage{subcaption}    
\setcitestyle{sort&compress}

\usepackage{booktabs}      
\usepackage{multirow}      
\usepackage{siunitx}       

\usepackage{enumitem}

\usepackage{xspace}

\AtBeginDocument{%
  }

\begin{document}

\title[KVServe]{KVServe: Service-Aware KV Cache Compression for Communication-Efficient Disaggregated LLM Serving}
\author{%
\normalsize
Zedong Liu$^{1,2,*}$,
Xinyang Ma$^{1,2,*}$,
Dejun Luo$^{1}$,
Hairui Zhao$^{2}$,
Bing Lu$^{2}$,
Wenjing Huang$^{2}$,
Yida Gu$^{2}$,
Xingchen Liu$^{2}$,\\
Zheng Wei$^{2}$,
Jinyang Liu$^{3}$,
Dingwen Tao$^{2}$,
Guangming Tan$^{2}$\\[0.35em]
$^{1}$University of Chinese Academy of Sciences \quad
$^{2}$Institute of Computing Technology, Chinese Academy of Sciences \quad \\
$^{3}$Shanghai Jiao Tong University
}

\renewcommand\footnotetextcopyrightpermission[1]{} 
\setcopyright{none}
\settopmatter{printacmref=false, printccs=false, printfolios=true}

\begin{abstract}
LLMs are widely adopted in production, pushing inference systems to their limits. Disaggregated LLM serving (e.g., PD separation and KV state disaggregation) improves scalability and cost efficiency, but it also turns KV into an explicit payload crossing network and storage boundaries, making KV  a dominant end-to-end bottleneck. Existing KV compression are typically static runtime configurations, despite production service context varies over time in workload mix, bandwidth, and SLO/quality budgets. As a result, a fixed choice can be suboptimal or even increase latency. We present \emph{KVServe}, the first service-aware and adaptive KV communication compression framework for disaggregated LLM serving: KVServe (1) unifies KV compression into a modular strategy space with new components and cross-method recomposition; (2) introduces Bayesian Profiling Engine that efficiently searches this space and distills a 3D Pareto candidate set, reducing  $50\times$ offline search overhead; and (3) deploys a Service-Aware Online Controller that combines an analytical latency model with a lightweight bandit to select profiles under constraints and correct offline-to-online mismatch. Integrated into vLLM and evaluated across datasets, models, GPUs and networks, KVServe\footnote{\url{https://github.com/hpdps-group/KVServe}} achieves up to $9.13\times$ JCT speedup in PD-separated serving and up to $32.8\times$ TTFT reduction in KV-disaggregated serving.

\end{abstract}



\maketitle

\renewcommand{\shortauthors}{Liu and Ma et al.}
\begingroup
\renewcommand{\thefootnote}{\fnsymbol{footnote}}
\footnotetext[1]{Equal contribution.}
\endgroup

\section{Introduction}

Large language models (LLMs) are becoming a general-purpose engine for production inference, yet their autoregressive generation requires maintaining and repeatedly accessing the \emph{Key Value (KV) cache} throughout decoding. In practice, LLM inference is commonly divided into two stages: \emph{prefill} and \emph{decode}. Prefill computes prompt KV cache in parallel and is typically compute-intensive.  Decode iteratively generates tokens and reads KV, making it more memory-intensive~\cite{zhou2024llmsurvey}.

To boost throughput and support long contexts at lower cost, production serving systems are moving to \emph{disaggregated} inference architectures. Two representative designs are \emph{prefill/decode (PD) separation} and \emph{KV state disaggregation}~\cite{zhong2024distserve,patel2024splitwise,qin2025mooncake}. In PD separation, prefill and decode run on separate GPU nodes to reduce co-location contention and to enable stage-specific scaling. In KV state disaggregation, the KV cache is offloaded to a storage hierarchy or remote KV pool to support longer contexts and cross-request reuse (e.g., RAG, and agents). Unlike monolithic serving where KV is internal GPU state, disaggregation makes KV an explicit payload that must be red across networks~\cite{zhang2025hack}.  As contexts grow, KV quickly becomes massive (eg. Llama 3.1-70B generates \emph{39.06 GB} KV at 128K tokens~\cite{schmid2024llama31}).

\begin{figure}[tbp] 
  \centering
  \includegraphics[width=\linewidth]{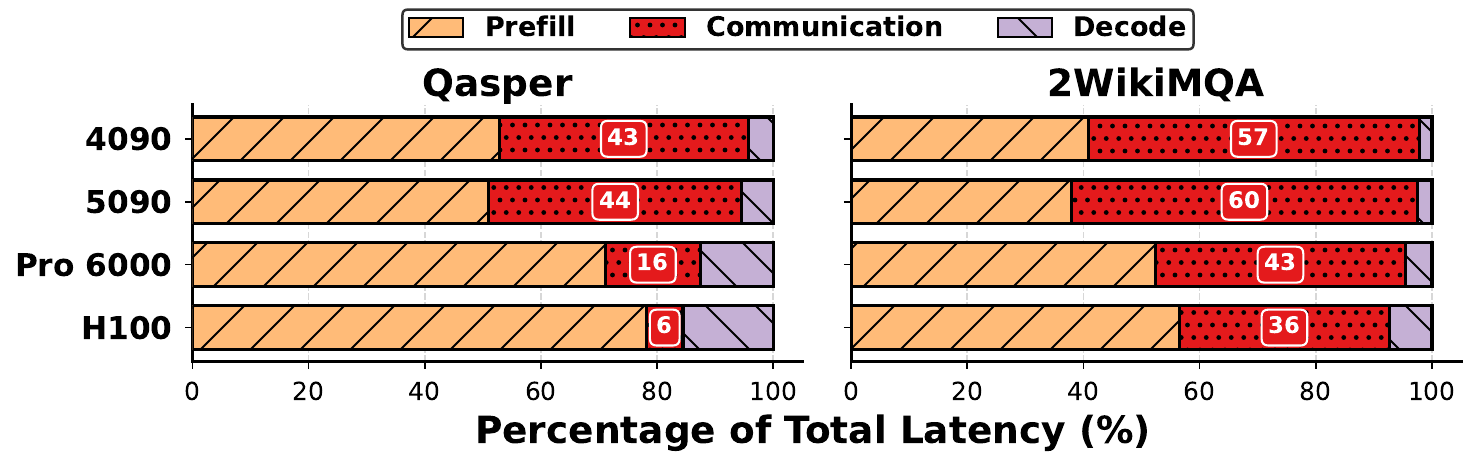}
  \caption{Time breakdown under PD-separated serving.}
  \label{fig:2}
\end{figure}

However, this disaggregation introduces a bandwidth-dependent bottleneck: the cost of transferring \emph{KV cache} across network/IO boundaries. 
Recent agentic and long-context workloads further amplify this pressure: their long inputs and short outputs allow prefill workers to generate KV cache at very high throughput. For example, serving 32K-token requests with Qwen3-235B on a 64-node prefill cluster requires 2.1 Tbps of KV egress bandwidth~\cite{qin2026prefill}.
In common cloud deployments, cross-cluster bandwidth is often constrained to below 100 Gbps.~\cite{aws_ec2_faq}. Similar limits apply to remote storage/KV pools, where throughput is often below 10 Gbps~\cite{Cachegen}. This makes KV  a dominant cost in disaggregated serving. In our end-to-end experiments (Fig.~\ref{fig:2}), KV communication time accounts for up to \emph{60\%} of job completion time. As KV cache grows, this bottleneck will further intensify, calling for optimizations. 

Recent work has proposed a range of \emph{KV compression} methods that significantly reduce KV volume with acceptable quality loss. Representative works such as CacheGen ~\cite{Cachegen}, KIVI ~\cite{KIVI}, and KVQuant ~\cite{KVQuant} quantize BF16 KV caches to 4-bit or 2-bit and further increase compression ratios via lossless coding. Finer-grained quantization schemes, such as mixed-precision quantization ~\cite{tao2025asymkv,liu2025pm,duanmu2024skvq}, assign different precisions based on layer-level or token-level importance. Other methods improve compressibility and control quality degradation through transforms such as Hadamard ~\cite{QuaRot} or Affine ~\cite{AffineQuant} preprocessing. 

Despite their effectiveness, these methods are generally \emph{statically configured} at runtime: fixed choice of transforms, quantization granularities, and codecs. A static configuration may reduce latency under some conditions, but can also cause \emph{negative optimization}. This is because the service context in production changes dynamically, including workload type, effective bandwidth, and Service Level Objective (SLO) budgets. Our measurements show that the latency-optimal choice can switch across workloads and bandwidth regimes (detailed in Sec.\ref{sec:2.3}). In other words, in disaggregated serving, KV compression is not a fixed algorithm choice; it is a \emph{constrained, service-state-dependent strategy selection} problem.

However, achieving service-aware and adaptive KV compression in disaggregated inference is non-trivial and faces three key challenges. First, existing KV compression methods are implemented as tightly coupled designs with incompatible code and parameter interfaces, making them difficult to reuse and compose into a plug-and-play interface. Second, abstracting KV compression into a searchable strategy space leads to an exponentially growing strategy space, making exhaustive profiling impractical. Third, online serving must meet quality and SLO budgets~\cite{qin2025mooncake}; selecting strategies based solely on compression ratio or quality can be infeasible or suboptimal, and there is a lack of a constrained theoretical model to guide online selection and switching.

To address these challenges, we present \emph{KVServe}. To the best of our knowledge, KVServe is the first \emph{service-aware} and \emph{adaptive} KV compression framework for disaggregated LLM serving. KVServe unifies KV compression techniques into a composable and extensible strategy space, senses online service context, and selects an optimal profile under quality and SLO constraints. Our key designs and contributions are:

\begin{figure}[t] 
  \centering
  \includegraphics[width=\linewidth]{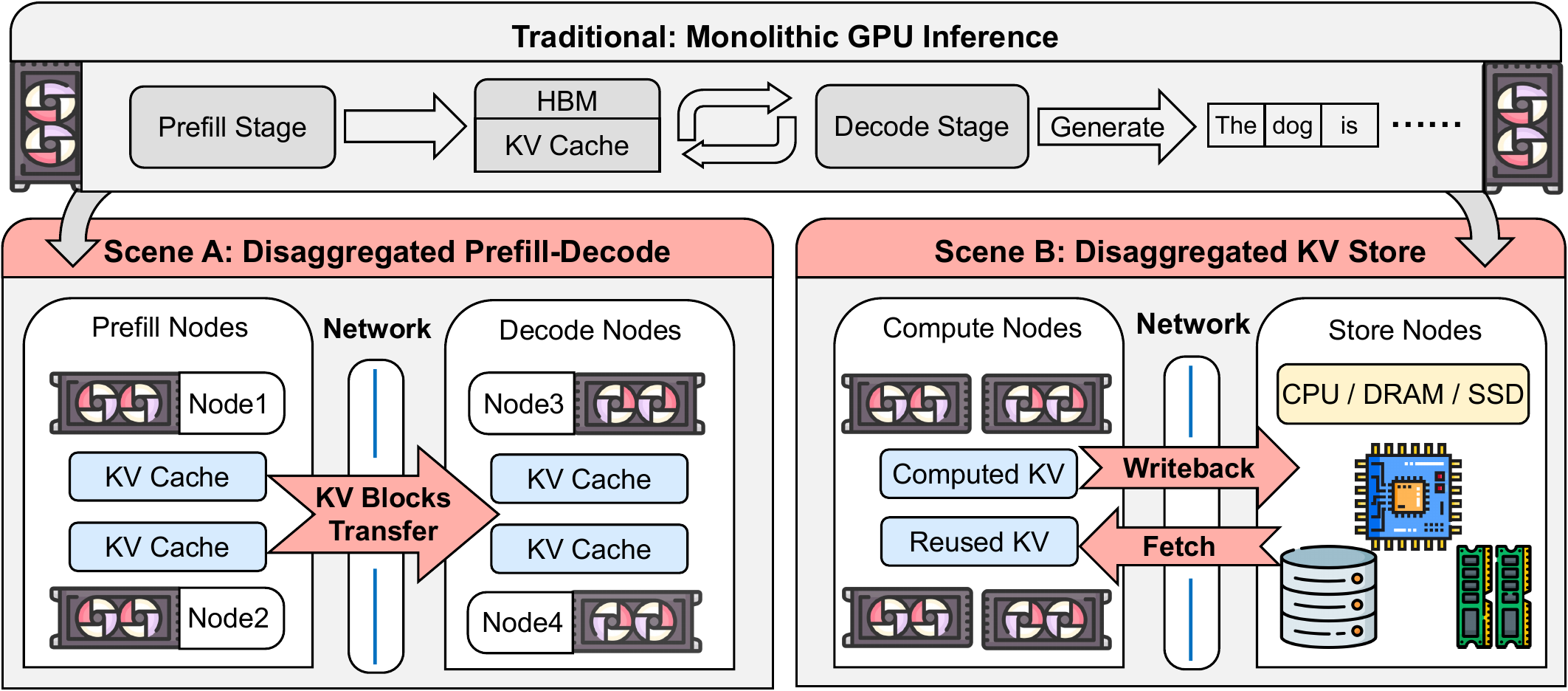}
  \caption{Architecture of disaggregated serving system. }
  \label{fig:1}
\end{figure}

\begin{itemize}[leftmargin=*]
\item We abstract KV compression as a unified modular pipeline and decompose representative methods into pluggable components. Building on this abstraction, we introduce a new quantization component designed by us; through cross-method composition and reuse, we form an enumerable and extensible strategy space.

\item We design an efficient \emph{Bayesian Profiling Engine}. Facing the combinatorial explosion of the strategy space, it uses Bayesian optimization to substantially reduce expensive end-to-end profiling runs, cutting offline search overhead from \emph{1000 hours} to the \emph{20-hour} scale.

\item We propose a \emph{Service-Aware Online Controller} that senses service context at runtime and rapidly selects the optimal profile from the offline candidates. The controller combines an analytical latency model with a lightweight bandit to correct mismatches between offline profiling and online execution, improving robustness to real-world drift.
\item We integrate KVServe into the vLLM inference pipeline and evaluate it across many datasets, models, and GPU/network configurations. Compared with the baseline and SOTA KV compression methods, KVServe achieves up to \emph{9.13}$\times$ JCT reduction in PD-separated serving, and up to \emph{32.8}$\times$ TTFT reduction in KV-disaggregated serving.
\end{itemize}

\section{Background and Motivation}
\subsection{Bottleneck in Disaggregated LLM Serving}

In recent years, the inference pressure of large language models has been driven by the dual scaling of \emph{model size} and \emph{context window}. Meanwhile, RAG and agentic workflows further push the demand for long-context online serving to accommodate more retrieved evidence and tool-call traces~\cite{arslan2024rag_survey,li2025agenticß}. 
Under this trend, production serving systems increasingly adopt \emph{disaggregated} architectures (Fig.~\ref{fig:1}), by separating compute and KV state across different nodes and remote storage pools~\cite{zhong2024distserve,patel2024splitwise,qin2025mooncake}. As a result, KV cache—previously resident in GPU memory—becomes an I/O payload that must be moved across devices over the network and moves onto the critical path of end-to-end latency.

\textbf{Compute disaggregation: Prefill/Decode separation.}
Prior work separates prefill and decode across GPU nodes to reduce co-location contention and scale each stage independently. Prefill produces the prompt KV cache and ships it to decode, which consumes the KV during generation, enabling stage-aware placement on heterogeneous GPU pools.
In practice, this split often breaks the shared high-speed interconnect domain (e.g., InfiniBand). With Ethernet-connected GPU nodes in the cloud, bandwidth limits can greatly amplify KV migration cost and make communication a dominant bottleneck.
We quantify this on Llama-3.1 with Qasper, using H100 decode and varying prefill instances: Fig.~\ref{fig:2} breaks down JCT into prefill, decode, and communication. At 10--50~Gbps, communication accounts for 16\%--60\% of JCT.

\textbf{State disaggregation: KV cache offloading and cross-query reuse.}
In RAG, multi-turn conversations, and templated requests, systems often exploit \emph{cross-query KV reuse} (e.g., prefix caching) to avoid redundant prefill, reducing TTFT and improving throughput. Keeping reusable KV resident in GPU memory is usually impractical: reuse can occur across requests far apart in time or on different GPU nodes, and GPU memory cannot hold many long-context KVs concurrently (often tens to hundreds of GB)~\cite{schmid2024llama31}. As a result, systems offload KV to CPU/SSD tiers or a remote KV pool, but remote reads become latency-critical. Under 5--15Gbps links in typical cloud servers, KV communication accounts for up to 66\% of end-to-end time~\cite{Cachegen}, making KV movement a key bottleneck for latency and SLO attainment.

\begin{figure}[t] 
  \centering
  \includegraphics[width=\linewidth]{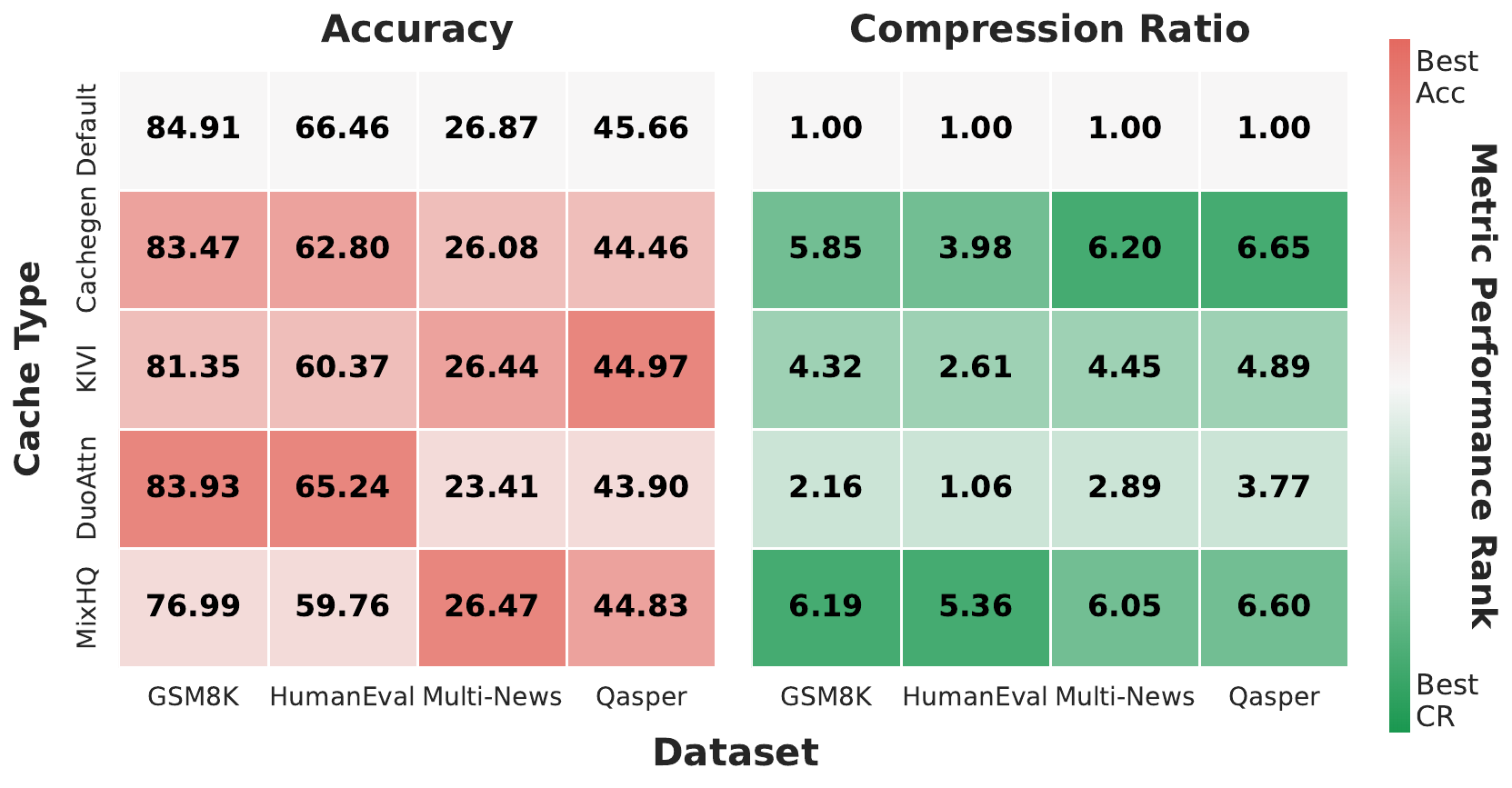}
  \caption{Accuracy and compression ratio across workloads .}
  \label{fig:performance_heatmap}
\end{figure}



\subsection{Rethinking KV Cache Compression: From Static to Service-Aware}
\label{sec:2.3}

In production LLM serving, requests are heterogeneous and are routinely \emph{typed} by workload (e.g., math reasoning, code generation, long-document QA) via task- and intent-aware routing at the ingress, so that different request types can be steered to appropriate backends or execution paths (e.g., industry routers such as Red Hat's \emph{LLM Semantic Router} and NVIDIA's \emph{LLM Router})~\cite{wang2025Router,nvidia_llm_router,ong2024routellm}. Accordingly, we treat the workload label $w$ for each session segment as a standard routing output of the serving stack (rather than a strong assumption), and focus on the service side: selecting a KV compression strategy conditioned on $w$ and online conditions. Crucially, different workload types often tolerate different levels of quality loss (i.e., different quality budgets), and the serving environment further evolves over time.

\textbf{Motivation 1: The Optimal KV Compression Strategy Varies Across Service Workloads. }
Existing KV compression methods are mostly \emph{statically configured}: e.g., using a fixed transform, a fixed quantization granularity, and a fixed codec. Such methods may achieve favorable compression ratio and accuracy on certain workloads, but their advantages do not generalize well across workloads. The reason is that different tasks exhibit substantially different request distributions and generation behaviors, which leads to systematic shifts in the statistics and compressibility of KV cache. As a result, the same compression strategy can yield markedly different accuracy and compression gains across tasks.
 
The results in Fig. ~\ref{fig:performance_heatmap} further validate this \emph{workload dependence}. For example, KIVI achieves the best accuracy on Qasper, but ranks near the bottom on GSM8K and HumanEval. In contrast, DuoAttention performs best on GSM8K and HumanEval, yet performs worst on Multi-News and Qasper. Similar instability appears not only in accuracy but also in compression ratio. CacheGen reaches the best compression ratio of 6.20$\times$ on Multi-News, but only 3.98$\times$ on HumanEval, which is lower than MixHQ's 5.36$\times$. These observations can be summarized as follows: a static KV compression strategy cannot be optimal across diverse workloads.

\begin{figure}[t] 
\captionsetup{belowskip=8pt} 
  \centering
  \includegraphics[width=\linewidth]{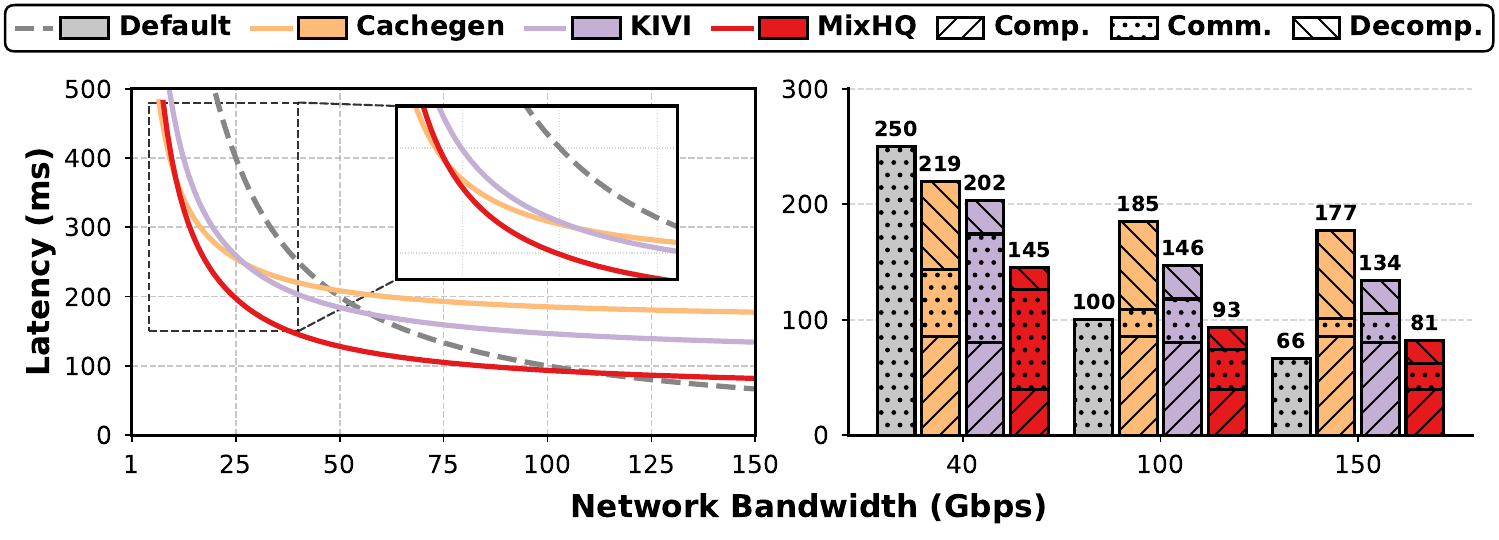}
  \caption{KV  latency across effective bandwidths (left) and  time breakdown (right).}
  \label{fig:_latency}
\end{figure}

\begin{takeawaybox}[TAKEAWAY-1]
There is no universally optimal KV compression strategy across workloads. Practical systems must reason over \emph{multiple candidate strategies} rather than committing to a single static configuration.
\end{takeawaybox}

\textbf{Motivation 2: The Optimal Strategy Also Depends on Bandwidth—and Can Even Hurt Performance. }
Beyond compression ratio, end-to-end speedup also depends on the service-side effective bandwidth and the compression/decompression throughput. For any compression strategy $p$, the KV  latency has two parts: (i) communication of the compressed KV and (ii) compression and decompression. Comparing to uncompressed  latency reveals speedup (or slowdown). Fig. ~\ref{fig:_latency} reports the KV  latency of CacheGen, MixHQ, and KIVI across bandwidths. The optimal strategy switches with bandwidth: CacheGen is optimal at very low bandwidth, but as bandwidth increases it is overtaken by MixHQ and then KIVI (two intersections), with MixHQ best over a broad range.

More importantly, each profile is beneficial only within a bandwidth regime: once bandwidth exceeds a threshold, communication savings no longer offset (de)compression, making latency worse than no compression. In Fig. ~\ref{fig:_latency}, the thresholds for the three methods are 50/55/110~Gbps, respectively. Therefore, if a system ignores bandwidth as a 
service state and applies a fixed static compression strategy, 
it cannot remain  optimal across network conditions and may even directly hurt performance in some cases.

\begin{takeawaybox}[TAKEAWAY-2]
The optimal KV compression strategy depends on dynamic service conditions (e.g., available bandwidth). A static, fixed strategy is unsafe in practice; KV compression must be \emph{adaptive} and \emph{service-aware} at runtime.
\end{takeawaybox}

\subsection{Challenges for Service-Aware KV Cache Compression}
\begin{figure}[t] 
  \centering
  \includegraphics[width=\linewidth]{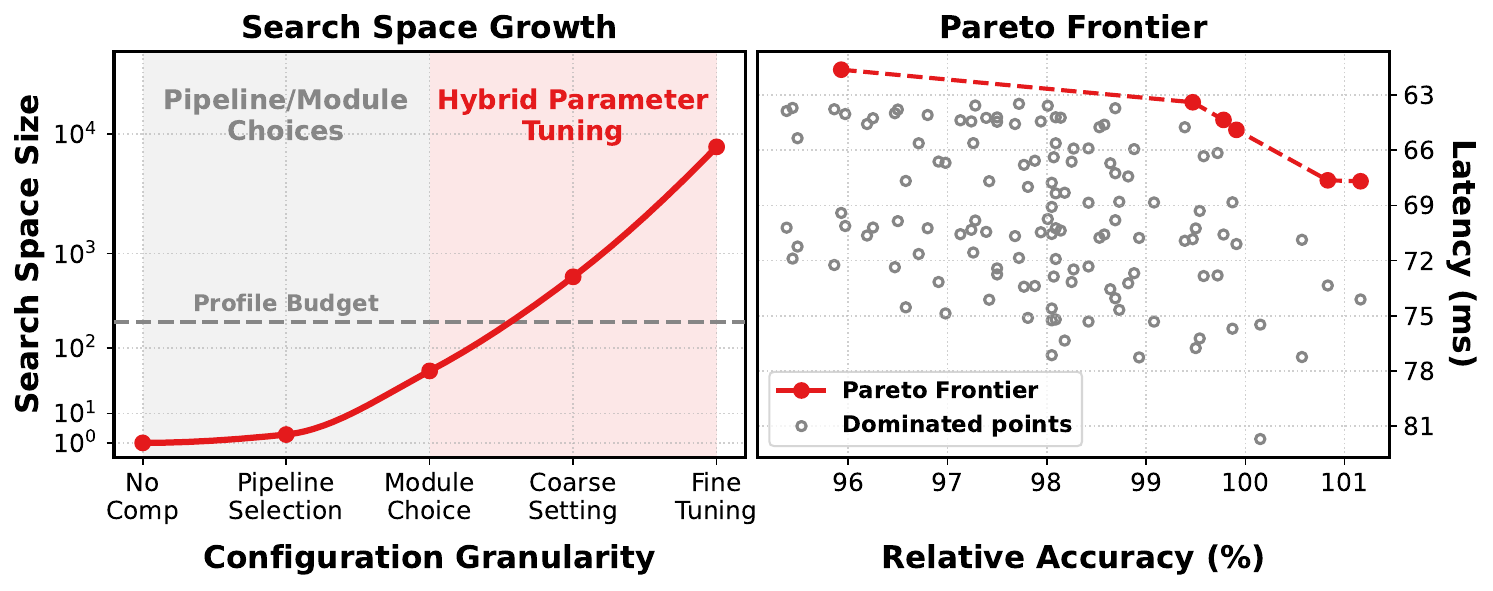}
  \caption{Left: Search space size under different granularities. Right: Latency–accuracy tradeoff of a collection of profiles from a representative pipeline.}
  \label{fig:challenge}
\end{figure}
\noindent\textbf{Challenge 1: The Combinatorial Explosion of the Strategy Space.} To address the limitations of static configurations revealed by Motivation 1, one can abstract KV compression as a searchable strategy space of components and parameters, and then select the best configuration offline for a target workload. The core challenge is combinatorial explosion: as we move from \emph{pipeline/module choices} to \emph{fine-grained parameter tuning}, the number of candidates grows roughly exponentially with the degrees of freedom. Fig. ~\ref{fig:challenge} (left) shows that enabling fine-grained tuning quickly expands the space to nearly $10^{4}$ candidates. Each candidate further requires an end-to-end profiling run (compression ratio, latency, and quality); in our setup this takes about 15 minutes, making exhaustive search cost tens to hundreds of GPU-hours—well beyond a practical offline budget. Therefore, our first challenge is to efficiently search this huge space while preserving candidate quality.

\noindent\textbf{Challenge 2: The Latency–Quality Tradeoff without a Clear Decision Principle.} Even after offline profiling compresses the space into a finite candidate set, online selection still faces an inherent latency--quality trade-off with no single metric that resolves it. Fig. ~\ref{fig:challenge} (right) plots 131 candidates under the same workload and shows a highly dispersed distribution: latency can differ markedly at similar quality levels, and further latency reductions often incur non-trivial quality loss. Hence, a production system must choose a feasible and optimal strategy under constraints such as SLO and an accuracy budget; ranking by compression ratio alone or quality alone can frequently yield infeasible or suboptimal profiles. This motivates a constrained model that jointly captures (de)compression overhead, post-compression  volume, and quality degradation, enabling interpretable selection and switching as service conditions change.

\section{Problem Formulation}

\subsection{Serving System Model}
\label{sec:serving_system_model}

We consider two common KV-movement paths in \emph{disaggregated LLM serving}: (i) prefill$\rightarrow$decode migration under \emph{PD separation}, and (ii) fetching/offloading KV under \emph{KV state offloading/reuse}. In both cases, KV becomes an explicit payload that crosses a network/IO boundary and contributes directly to end-to-end latency.
We therefore use a \emph{request} as the decision granularity: the system selects a compression profile when the request’s KV movement begins and keeps it consistent throughout the request. Crucially, the realized communication cost is governed by the \emph{effective} network/IO regime—application-level goodput under contention—rather than nominal link bandwidth. Accordingly, we incorporate lightweight runtime communication signals into the service context to enable network-aware, constraint-driven profile selection within each request.
The service context within this window is abstracted as:
{ \small
\[
c = (w, B, T_{\text{SLO}}, q_{\min}),
\]
}
where $w$ denotes the workload class of the session segment (provided by an upper-layer router/classifier; we do not study its implementation),
$B$ is the currently available \emph{effective bandwidth} (a unified abstraction of network or I/O goodput),
$T_{\text{SLO}}$ is the latency budget for the session segment, and $q_{\min}$ is the minimum quality requirement.

A KV compression strategy (profile) can be represented by a parameterized triple:
{ \small
\[
p = (cr_p, s_p, q_p),
\] }
where $cr_p$ is the compression ratio, defined as $cr_p \triangleq \frac{V}{V_p}$, with $V$ being the total amount of uncompressed KV to be moved within the session segment (in bytes)
and $V_p$ being the total compressed KV size under strategy $p$.
$s_p$ is the effective (de)compression throughput (bytes/s), defined as the harmonic mean of the encoding throughput $s_{p}^{\text{enc}}$
and the decoding throughput $s_{p}^{\text{dec}}$:
{ \small
\[
s_p \triangleq \left(\frac{1}{s_{p}^{\text{enc}}}+\frac{1}{s_{p}^{\text{dec}}}\right)^{-1}
= \frac{s_{p}^{\text{enc}}\, s_{p}^{\text{dec}}}{s_{p}^{\text{enc}}+s_{p}^{\text{dec}}},
\]
}
so that the total encoding and decoding time can be written as
$\frac{V}{s_{p}^{\text{enc}}} + \frac{V}{s_{p}^{\text{dec}}} = \frac{V}{s_p}$.
Finally, $q_p$ denotes the quality metric of strategy $p$ under workload $w$
(e.g., task accuracy or an equivalent measure of quality loss).

Given a dynamic service context $c$, our goal is to select a strategy $p$ for each session segment that satisfies
the service requirements $T_{\text{SLO}}$ and $q_{\min}$ while optimizing end-to-end performance;
the latency model and the resulting optimization problem are presented in the next section.

\subsection{Constrained Optimization}

Within each \emph{session segment} decision window, we use the segment-level end-to-end completion time, Job Completion Time (JCT), as the optimization target.
We decompose it into two parts: (i) the model execution cost that is independent of the KV compression strategy, and (ii) the additional cost introduced by KV (de)compression and KV movement.
Let $V$ denote the total amount of \emph{uncompressed} KV that must cross the boundary within the segment (in bytes), and let $B$ denote the \emph{effective bandwidth} (bytes/s) observed by KV movement during online serving.
Let $T_{\text{model}}(w)$ denote the model execution cost under workload class $w$, which is approximately invariant to the choice of compression strategy given a fixed model and serving configuration; we also absorb other strategy-independent operator execution and scheduling overheads into $T_{\text{model}}(w)$.

For any compression strategy $p=(cr_p,s_p,q_p)$, the compressed KV volume is $V_p = \frac{V}{cr_p}$.
Using the definition of the effective (de)compression throughput $s_p$ from ~\ref{sec:serving_system_model}, we model the segment JCT as:
\begin{equation}
\small
\label{eq:latency_model}
T_p(c)=T_{\text{model}}(w)+\frac{V}{s_p}+\frac{V}{B\,cr_p},\,
T_0(c)=T_{\text{model}}(w)+\frac{V}{B}.
\end{equation}
Here, $\frac{V}{s_p}$ represents the sum of encoding and decoding time.
We assume that the amount of data processed by (de)compression is of the same order as the KV volume to be moved, and we include operator execution and scheduling overheads unrelated to KV (de)compression in $T_{\text{model}}(w)$.

Online strategy selection under service context $c$ must satisfy the segment-level latency budget and the minimum quality requirement, and we select a profile to minimize $T_p(c)$ under these requirements.
For convenience, we define the feasible set of strategies under context $c$ as
\begin{equation}
\small
\mathcal{P}(c)\triangleq \left\{\,p\in\mathcal{P}\;\middle|\; T_p(c)\le T_{\text{SLO}},\; q_p(w)\ge q_{\min}\right\},
\end{equation}
where $\mathcal{P}$ is the set of selectable compression strategies.
We then formulate the segment-level strategy selection as the following constrained optimization problem:
\begin{equation}
\small
p^*(c)\in \arg\min_{p\in\mathcal{P}(c)} \; T_p(c).
\end{equation}

This formulation explicitly captures the joint effect of four factors: the effective bandwidth $B$ determines the upper bound of time savings from compression, the effective throughput $s_p$ determines the additional (de)compression overhead, the compression ratio $cr_p$ determines the red KV volume after compression, and $q_p(w)$ captures the quality cost.
In the following sections, we derive benefit conditions from this model and design a policy that selects and switches strategies in response to changing conditions.

\begin{figure}[t] 
  
  \centering
  \includegraphics[width=\linewidth]{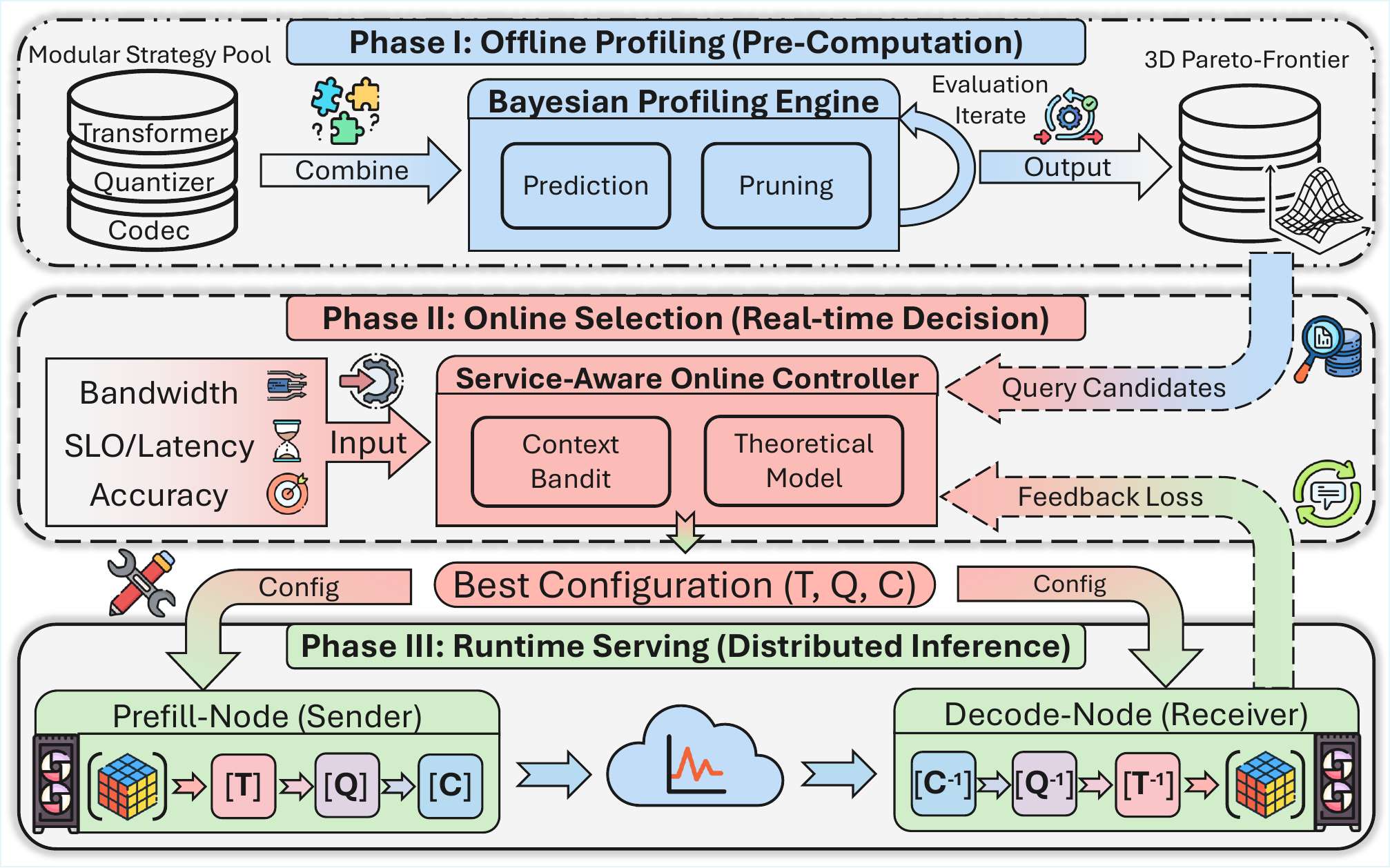}
  \caption{Overview Architecture of KVServe.}
  \label{fig:overview}
\end{figure}

\section{Design Overview}

To address the above problems and challenges, we propose KVServe. To the best of our knowledge, KVServe is the first \emph{service-aware} and \emph{adaptive} KV communication compression framework for \emph{disaggregated LLM serving}. Unlike prior approaches that rely on static configurations to optimize a single metric, KVServe unifies mainstream KV compression techniques into a composable and extensible strategy space, and adapts to online service conditions to select the optimal KV compression strategy. Under SLO and quality constraints, KVServe aims to minimize end-to-end latency. KVServe consists of three core components (shown in Fig.~\ref{fig:overview}):

\begin{itemize}[leftmargin=*]
\item \textbf{Modular Strategy Pool.} We abstract KV compression as a modular pipeline composed of pluggable components, and map representative existing methods into this abstraction. Beyond incorporating improved variants of existing components, we also enable new components to be designed and integrated, forming an enumerable space.

\item \textbf{Bayesian Profiling Engine.} Facing the combinatorial explosion of the strategy space, the profiling engine uses Bayesian Optimization with Gaussian Processes to substantially reduce the number of expensive end-to-end profiling runs. It ultimately derives a candidate set defined by a 3D Pareto frontier for fast online selection.

\item \textbf{Service-Aware Online Controller.} During online inference, the controller senses the service context and selects the optimal profile from the offline candidate set. It has two layers: (i) an analytical latency model that provides interpretable end-to-end benefit estimates and derives benefit boundaries; and (ii) a lightweight online bandit that refines decisions based on runtime observations, correcting system drift and improving robustness.
\end{itemize}

Overall, KVServe operates in three stages: \emph{Offline Profiling}, \emph{Online Selection}, and \emph{Runtime Serving}. In \emph{Offline Profiling}, the \emph{Bayesian Profiling Engine} efficiently searches the \emph{Modular Strategy Pool} and constructs the candidate set. In \emph{Online Selection}, the \emph{Service-Aware Online Controller} chooses the most suitable compression profile given the current service state and constraints. Finally, in \emph{Runtime Serving}, KVServe executes the selected strategy at KV movement boundaries. 

\begin{figure}[b] 
  \centering
  \includegraphics[width=\linewidth]{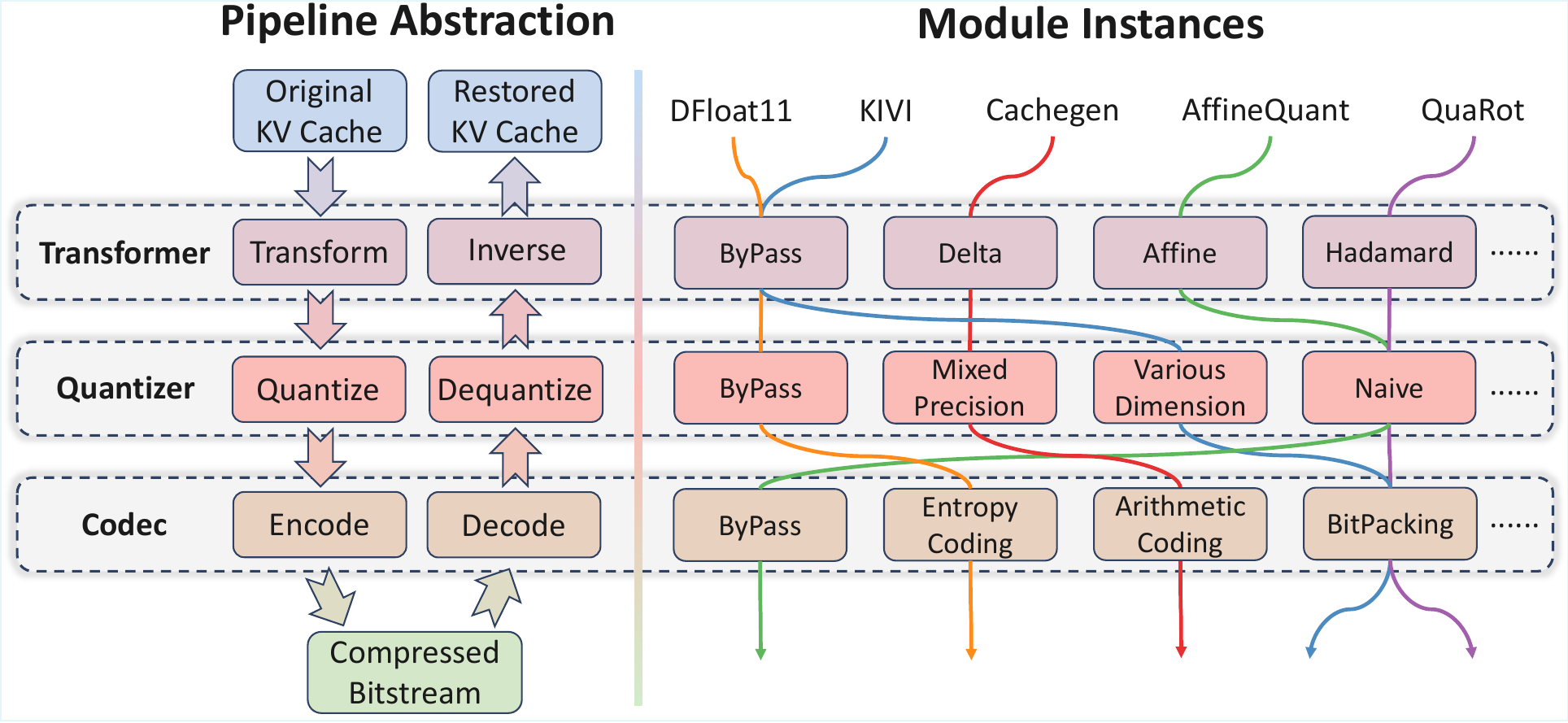}
  \caption{The Unified KV Cache Compression Pipeline.}
  \label{fig:compression_pipeline}
\end{figure}

\section{Offline Profiling Engine}
\label{sec:offline_engine}

\subsection{Constructing the strategy space}
\label{sec:config_space}

Existing KV cache optimizations—including rotation~\cite{QuaRot}, quantization~\cite{KIVI}, and entropy coding~\cite{DFloat11}—are predominantly studied in isolation, often yielding suboptimal trade-offs between compression ratio (CR) and accuracy (Acc). To bridge this gap, we propose a generalized \textit{KV Cache Compression Pipeline} that unifies these disjoint strategies into a composable framework, recasting compression as a search problem over a comprehensive strategy space.

\noindent\textbf{Pipeline Abstraction and Module Instantiation.}
We formalize the KV cache compression lifecycle as a sequential composition of three distinct stages, $\mathbf{BS} = \mathcal{C}\left(\mathcal{Q}\left(\mathcal{T}(\mathbf{X})\right)\right)$, as schematically illustrated in Fig.~\ref{fig:compression_pipeline}: 

\noindent\cnum{1} \textbf{Transformer ($\mathcal{T}$):} A pre-processing stage reshaping distributions to facilitate downstream compression. Modules include Delta~\cite{Cachegen}, Hadamard ~\cite{QuaRot} and Affine ~\cite{AffineQuant}.
    
\noindent\cnum{2} \textbf{Quantizer ($\mathcal{Q}$):} The primary stage for bit-width reduction. This module encompasses multi-dimensional quantization methods ~\cite{KIVI} and supports \textit{Mixed-Precision Quantization} at both layer-wise and head-wise granularities.
    
\noindent\cnum{3} \textbf{Codec ($\mathcal{C}$):} The final stage encodes the data stream to minimize footprint. We integrate the high-performance library nvCOMP ~\cite{nvCOMP} library to support efficient algorithms.

By decomposing existing SOTA methods into these atomic components, KVServe enables the exploration of their Cartesian product. This extensible architecture allows for arbitrary combinations (e.g., pairing a QuaRot transformer with a CacheGen quantizer) to identify synergistic configurations that outperform isolated baselines.

\noindent\textbf{Mixed-Precision Head-Wise Quantization (MixHQ).}
In this pipeline, we also propose a novel framework for $\mathcal{Q}$ that shifts the paradigm from binary pruning~\cite{DuoAttention} to variable precision allocation. By distinguishing between Retrieval Heads and Streaming Heads, MixHQ applies aggressive ultra-low bit-width quantization to the latter instead of discarding them, while retaining Retrieval Heads in high precision to preserve critical long-range dependencies.

Crucially, this framework is orthogonal to the granularity of importance estimation. It supports seamless generalization to the layer dimension (assigning lower bit-widths to deeper layers like PyramidKV ~\cite{PyramidKV}) and the token dimension (preserving heavy-hitters like SnapKV ~\cite{SnapKV}). This flexibility enables integration with various importance scoring methods, effectively transforming discrete pruning decisions into a continuous spectrum of precision allocation.

\subsection{Bayesian Profiling Engine}
\label{sec:search_methodology}
\subsubsection{Profiling Analysis and Optimization Strategy}
\label{sec:bo_gp}

To identify the optimal pipeline balancing CR and Acc, we must navigate a massive combinatorial strategy space $\mathcal{S}$.
As illustrated in \emph{Motivation 1} (Fig.~\ref{fig:challenge} left), the search space grows exponentially as configuration granularity deepens from \textit{Pipeline/Module Choices} to \textit{Hybrid Parameter Tuning}.
This explosive complexity renders brute-force methods impractical, necessitating a highly automated search strategy.

\begin{figure}[tbp]
  \centering
  \includegraphics[width=\linewidth]{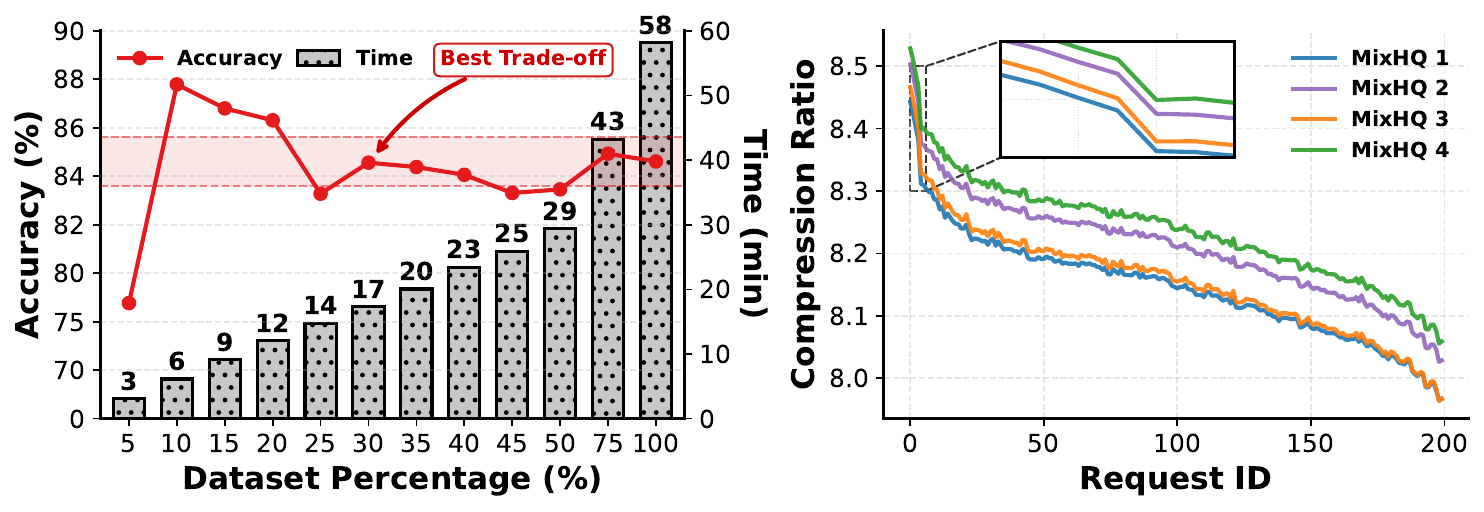}
  \caption{Profiling Efficiency and Ranking Consistency.}
  \label{fig:offline_motivation}
\end{figure}

Through empirical analysis, we derive two critical observations guiding our engine design:

\noindent\textbf{Observation 1: High Cost of Acc Evaluation.} 
Fig.~\ref{fig:offline_motivation} (left) reveals that executing full-dataset inference is prohibitively expensive. However, accuracy on uniformly sampled subsets stabilizes quickly, approximating full performance with negligible error. Thus, we employ sampled data as a reliable proxy to accelerate profiling.

\noindent\textbf{Observation 2: Stability of CR Relative Rankings.} 
Although absolute compression ratios fluctuate with content, Fig.~\ref{fig:offline_motivation} (right) demonstrates that the relative ranking of configurations remains strictly invariant across requests, even among MixHQ candidates with highly proximate ratios. This stability ensures that high-performing configurations identified offline reliably translate to online optimality.

Guided by these insights, we formulate the task as a \textit{Constrained Black-Box Optimization} problem. We adopt \textit{Bayesian Optimization (BO) with Gaussian Processes (GP)} over evolutionary algorithms or random search for two reasons: (i) \textit{sample efficiency}---given the high evaluation cost, minimizing the number of iterations is crucial, and BO leverages a surrogate model to approach the global optimum with far fewer samples; and (ii) \textit{uncertainty modeling}---GPs estimate both the mean and variance, enabling principled \textit{exploration--exploitation} trade-offs and reducing the risk of getting trapped in local optima.

Formally, given a configuration $\mathbf{c} \in \mathcal{S}$, we solve:
{\small
\[
    \max_{\mathbf{c}} \text{CR}(\mathbf{c}) \quad \text{s.t.} \quad \text{Acc}(\mathbf{c}) \ge \text{Acc}_{\text{threshold}},
\]
}
where $\text{CR}(\mathbf{c})$ and $\text{Acc}(\mathbf{c})$ denote the compression ratio and model accuracy, respectively.

\begin{algorithm}[t]
\caption{Constraint-Aware Bayesian Optimization with Gaussian Processes}
\label{alg:bo_gp}
\footnotesize
\SetAlgoVlined
\DontPrintSemicolon
\KwIn{
    \begin{tabular}[t]{@{}l@{}}
    Strategy Space $\mathcal{S}$; Accuracy Thres $Acc_{ths}$; \\
    Pruning Buffer $\epsilon$; Max Iterations $T_{max}$;
    \end{tabular}
}
\KwOut{Feasible configuration set $\mathcal{F}$;}

$\mathcal{S}_{emb} \leftarrow \text{OneHot}(\mathcal{S}_{cat}) \cup \text{MinMax}(\mathcal{S}_{num})$\;

Initialize GP Model $\mathcal{M}_{GP}$ and Observation Set $\mathcal{D}$\;

\For{$t \leftarrow 1$ \KwTo $T_{max}$}{
    Fit $\mathcal{M}_{GP}$ on $\mathcal{D}$\;
    $\lambda \leftarrow \text{GetExplorationWeight}(t)$\;
    
    
    $c_{curr} \leftarrow \operatorname*{argmax}_{c \in \mathcal{S}_{emb}} \text{AF}(\mathcal{M}_{GP}, c, \lambda)$\;
    
    $Acc_{curr}, CR_{curr} \leftarrow \text{Evaluate}(c_{curr})$\;
    $\mathcal{D} \leftarrow \mathcal{D} \cup \{(c_{curr}, CR_{curr}, Acc_{curr})\}$\;

    \If{$Acc_{curr} \ge Acc_{ths}$}{
        $\mathcal{S}_{emb} \leftarrow \mathcal{S}_{emb} \setminus \{c \mid CR(c) < CR_{curr} - \epsilon\}$\;
        $\mathcal{F} \leftarrow \mathcal{F} \cup \{c_{curr}\}$\;

    }
    \ElseIf{$Acc_{cur} \ll Acc_{ths}$}{
        $\mathcal{S}_{emb} \leftarrow \mathcal{S}_{emb} \setminus \{c \mid CR(c) > CR_{curr} + \epsilon\}$\;
    }

    $k_{fail} \leftarrow \text{UpdateFailureTimes}(Acc_{curr}, Acc_{ths}, t)$\;
    
    \If{$\text{CheckEarlyStopping}(\mathcal{S}_{emb}, \mathcal{D}, k_{fail})$}{
        \textbf{break}\;
    }
}

\Return $\mathcal{F}$\;
\end{algorithm}

\begin{figure}[t] 
  \centering
  \includegraphics[width=\linewidth]{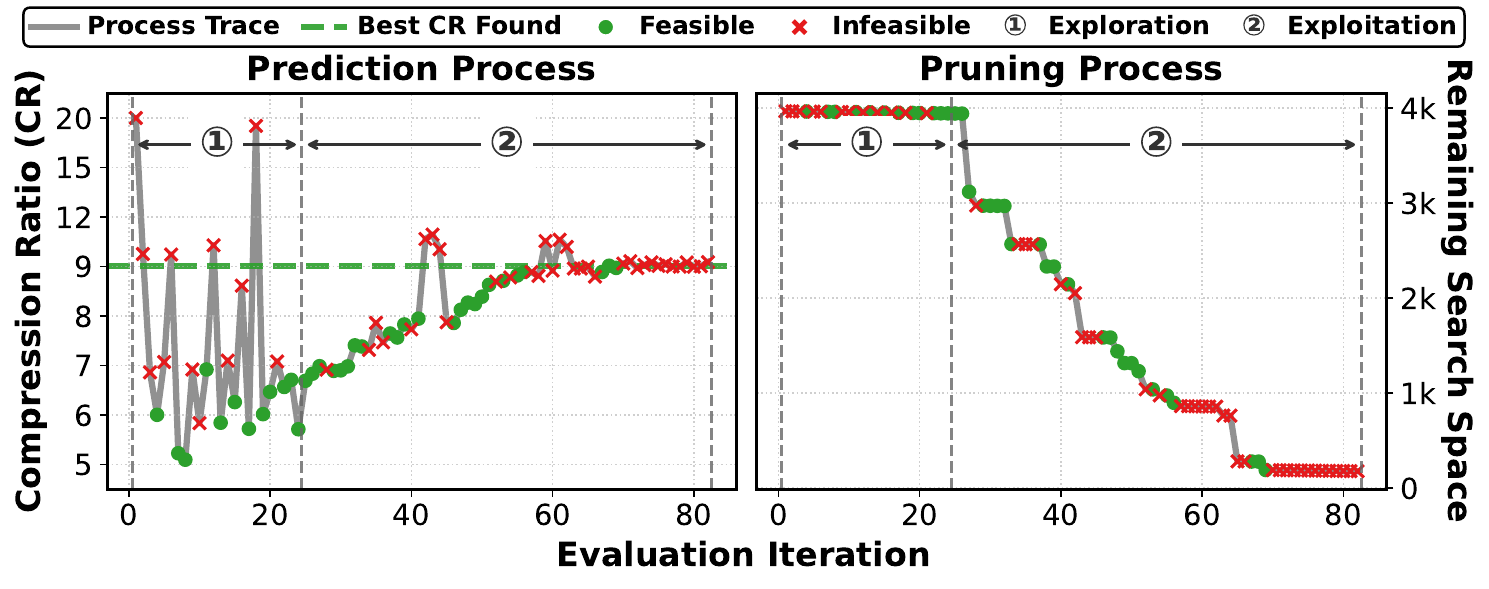}
  \caption{Prediction and Pruning Process Visualization.}
  \label{fig:search_process}
\end{figure}

\subsubsection{Bayesian Optimization with Gaussian Process}
\label{sec:offline_optimization}

Our profiling engine operates on a Bayesian Optimization cycle. It iteratively models the configuration-to-accuracy mapping using a Gaussian Process Surrogate Model and selects the next candidate to evaluate based on a utility score, repeating this process until convergence or early stopping.

\textbf{Acquisition Function (AF).} To guide the selection, we design a custom utility $\alpha(\mathbf{c})$ that balances maximizing the expected CR within constraints (\textit{Exploitation}) against reducing uncertainty (\textit{Exploration}):
\begin{equation}
\small
\label{eq:acquisition_func}
    \alpha(\mathbf{c}) = \underbrace{\text{CR}(\mathbf{c}) \cdot P(\text{Feasible})}_{\text{Exploitation}} + \underbrace{\lambda_{t} \cdot \sigma_{norm}(\mathbf{c})}_{\text{Exploration}},
\end{equation}
where $P(\text{Feasible})$ is the probability of satisfying the accuracy constraint derived from the GP posterior, and $\lambda_{t}$ is an exploration weight that decays over iterations $t$.

However, generic BO is ill-suited for our heterogeneous search space (mixed categorical/continuous parameters) and strict offline time constraints. Unlike standard asymptotic convergence, we require pinpointing optimal configurations within extremely limited iterations. Consequently, we enhance the architecture with specific optimizations for \textit{prediction} and \textit{pruning}, as detailed in Alg.~\ref{alg:bo_gp}.

\begin{figure}[t]
    \centering
    \includegraphics[width=0.9\linewidth]{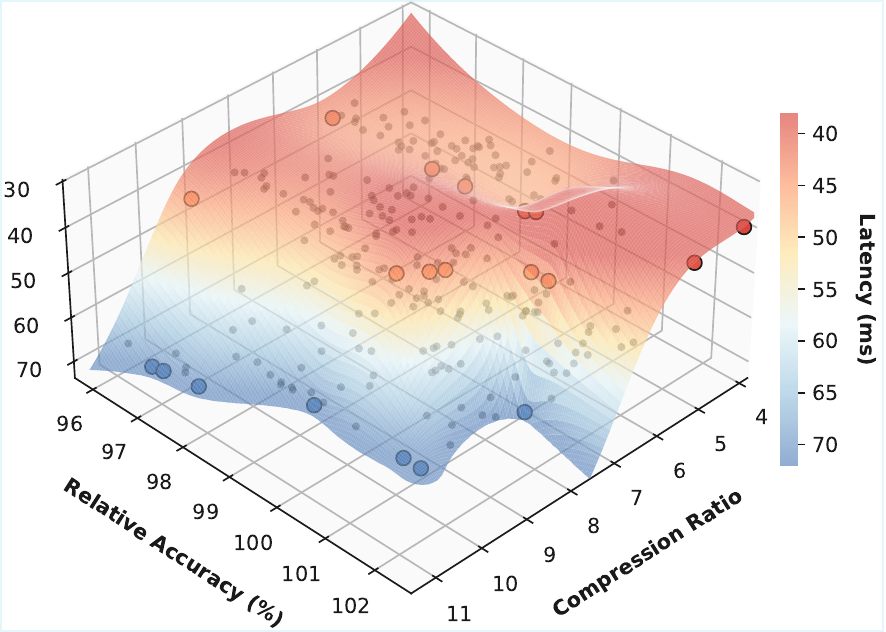} 
    \caption{The 3D Pareto Frontier of the Strategy Spaces. }
    \label{fig:3d_pareto}
\end{figure}

\noindent\textbf{Heterogeneous-Parameter Encoding (Line 1).} 
To resolve metric incompatibility in our mixed-parameter space, we map categorical and numerical variables to a unified embedding $\mathcal{S}_{emb}$ via One-Hot and Min-Max scaling, ensuring the GP kernel correctly measures structural similarity.

\noindent\textbf{Exploration-Exploitation Strategy (Lines 5-6).} 
We employ a dynamic strategy where the exploration weight $\lambda_t$ decays exponentially. This transitions the search from global exploration (high uncertainty sampling) to rapid exploitation (convergence on optima), as visualized in Fig.~\ref{fig:search_process} (left).

\noindent\textbf{Bi-Directional Pruning (Lines 9-13).} 
Leveraging the monotonic CR-Acc trade-off, we prune bi-directionally during exploitationas, as shown in Fig.~\ref{fig:search_process} (right): discarding higher-CR candidates if infeasible, and lower-CR ones if feasible, focusing solely on maximizing compression.

\noindent\textbf{Early-Stopping Mechanism (Lines 14-16).} 
To minimize overhead, the engine terminates execution early if consecutive failures $k_{fail}$ exceed a pre-defined limit or if the effective search space is exhausted.

The efficacy of these strategies is visualized in Fig.~\ref{fig:search_process}. While an exhaustive search of over 4,000 candidates would require $\sim$1,000 hours, our algorithm converges in fewer than 80 iterations (\textit{$\sim$20 hours}). This represents a \textit{50}$\times$ reduction in profiling overhead, effectively transforming an intractable exponential search into a manageable offline task.

\subsubsection{Outcome: The 3D Pareto Frontier}
\label{sec:outcome_pareto}

The output $\mathcal{F}$ from Alg.~\ref{alg:bo_gp} contains all feasible history, yet many configurations are dominated. Furthermore, maximizing CR alone is insufficient in networked serving, as computational overhead can negate communication gains. To address this, we introduce \textit{Latency} as a third critical dimension for evaluation.

We compute the \textit{3D Pareto Frontier} by projecting $\mathcal{F}$ into Acc-CR-Lat space, retaining only non-dominated points. As shown in Fig.~\ref{fig:3d_pareto}, the resulting surface represents optimal trade-offs among quality, footprint, and delay.

This \textit{3D Pareto Frontier} serves as a static runtime lookup table. It provides a candidate set for the \textit{Online Selection} (Sec. ~\ref{sec:online_controller}) to choose optimal strategies under dynamic context like bandwidth and SLO constraint.

\section{Service-Aware Online Controller}
\label{sec:online_controller}

Using the profiling engine in Sec. ~\ref{sec:offline_engine}, we shrink the massive strategy space into a finite 3D Pareto candidate set. However, a candidate set alone is insufficient in production. The system must sense online context (e.g., bandwidth, SLO, and quality budget) and select the latency-minimizing compression strategy with negligible overhead, while remaining robust to offline-to-online drift. To this end, we introduce a \emph{Service-Aware Online Controller}. The controller is built on an interpretable analytical latency model and further corrects runtime perturbations via a lightweight learnable bandit.

\subsection{Analytical Model}
\label{sec:controller_analytic}

In disaggregated LLM serving, KV movement occurs at clear system boundaries, such as prefill$\rightarrow$decode migration in PD separation or fetching from a remote KV pool. We use a \emph{request} as the decision granularity: the system selects a profile $p$ at the start of KV movement and keeps it fixed for the request. Given context $c=(w,B,T_{\text{SLO}},q_{\min})$, and under a workload $w$ the request JCT follows Eq.~\eqref{eq:latency_model}, subject to latency and quality budgets.

To ensure that the chosen profile meets the quality requirement, we bucket profiles by accuracy loss and restrict selection to the bucket matching the request's quality budget. After fixing a quality bucket $b$, the key variables for online selection reduce to each profile's compression ratio $cr_p$ and effective throughput $s_p$. We first ask a fundamental question: \emph{when does compression actually yield end-to-end speedup?} Using the latency model in Sec.~\ref{sec:serving_system_model}, by comparing $T_p(c)$ with $T_0(c)$, we can express a benefit condition: $T_0(c)/T_p(c) > 1$. This leads to a bandwidth-threshold condition in Eq.~\eqref{eq:benefit_threshold}.
\begin{equation}
\small
\label{eq:benefit_threshold}
B_p^\star \triangleq \left(1-\frac{1}{cr_p}\right)s_p,
\qquad
T_p(c) < T_0(c)\ \Longleftrightarrow\ B < B_p^\star.
\end{equation}

Notably, we observe that the condition is independent of the KV volume $V$ and depends only on the compression ratio and (de)compression throughput; moreover, the condition collapses to a threshold on the effective bandwidth $B$. This yields our first theorem.

\begin{theorem}[Benefit condition: bandwidth threshold]
\label{thm:benefit}
For any profile $p$, its offline parameters (e.g., $cr_p$ and $s_p$) determine a bandwidth threshold $B_p^\star$ (Eq.~\eqref{eq:benefit_threshold}). The profile is beneficial if $B < B_p^\star$; otherwise it is non-beneficial and can be filtered online, substantially shrinking the candidate set.
\end{theorem}

After filtering, we further use the latency model to characterize \emph{which profile is optimal under a given bandwidth}. For a workload $w$ and quality bucket $b$, we minimize $T_p$ over the feasible set $\mathcal{P}_b(w)$. For analysis, we let $x = 1/B$ and rewrite $T_p$ as a linear function of $x$,
\begin{equation}
\small
\label{eq:Tp_linear}
\tilde{T}_p(x) = \frac{T_p(c)-T_{\text{model}}(w)}{V}
= \frac{1}{s_p} + \frac{1}{cr_p}\,x,
\qquad x = \frac{1}{B}.
\end{equation}

This yields the following structural result.

\begin{theorem}[Piecewise-optimal policy]
\label{thm:piecewise}
For a workload $w$ and quality bucket $b$, minimizing $\tilde{T}_p(x)$ over $p\in\mathcal{P}_b(w)$ is equivalent to taking the lower envelope of the lines $\{\tilde{T}_p(x)\}$. Hence the optimal profile is piecewise constant in $x=1/B$: there exist breakpoints $0=x_0<x_1<\cdots<x_m$ such that for any $x\in[x_i,x_{i+1})$, the optimal profile is $p_i\in\mathcal{P}_b(w)$.
\end{theorem}

Together, Theorems~\ref{thm:benefit} and~\ref{thm:piecewise} provide an efficient and interpretable baseline selection mechanism. Offline, we construct the lower envelope in each quality bucket and obtain a piecewise policy table. Online, given the measured bandwidth $B$, we first apply Theorem~\ref{thm:benefit} to filter obviously non-beneficial profiles, yielding a tighter candidate set. Then, by Theorem~\ref{thm:piecewise}, we only need to look up the interval for $x=1/B$ to return the optimal profile $p_i$, and simultaneously return the neighboring profiles as a candidate set. This analytical mechanism achieves $O(1)$ decision cost, but it may still be affected by online drift. Next, we introduce lightweight online learning to perform residual correction for the mismatch between offline profiling and real serving conditions.

\subsection{Residual-Corrected Bandit}
\label{sec:controller_bandit}

\begin{figure}[tbp]
    \centering
    \includegraphics[width=\linewidth]{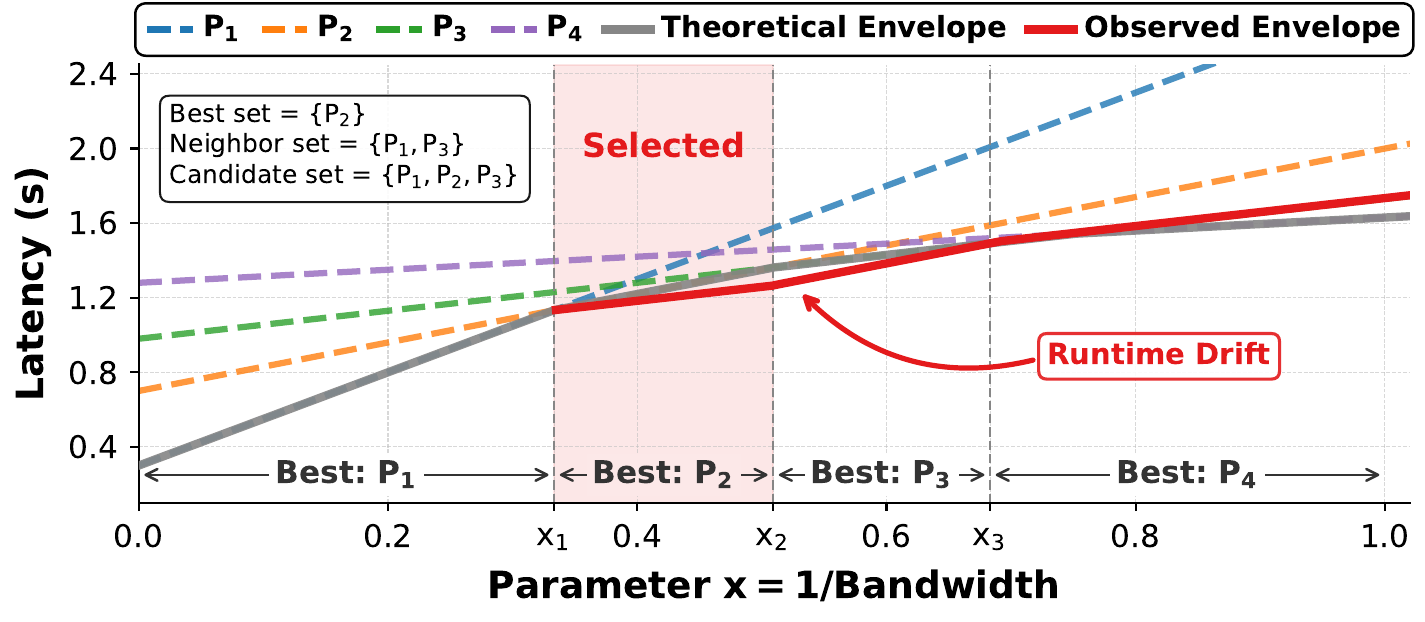} 
    \caption{Candidate set generation and bandit-based residual correction on the lower envelope.}
    \label{fig:lower_envelope}
\end{figure}

In online serving, parameters estimated from offline profiling often drift from reality. For example, GPU load and queue contention change the actual (de)compression throughput, and system scheduling and concurrency introduce additional overhead. As a result, the analytical model's latency predictions can deviate from runtime observations. Relying solely on offline parameters may cause the system to deviate from the true optimum during certain periods and require frequent re-profiling or manual retuning.

To address this, we add an extremely lightweight online learning layer on top of the analytical model to perform residual correction (Fig.~\ref{fig:lower_envelope}). The key idea is that the analytical model provides a strong prior by proposing the best profiles for the current bandwidth interval, and the online bandit only learns the difference between model prediction and runtime observation, achieving robustness at low cost.

Theorem~\ref{thm:piecewise} shows that the optimal policy is piecewise constant in $x=1/B$; under mild online drift, the most likely change is that the optimal choice switches among adjacent segments of the lower envelope. Therefore, we construct a tiny candidate set centered on the model-optimal profile $p^{\text{model}}_{b,i}$ for bucket $b$ and interval $i$, augmented by $1$--$2$ neighboring profiles on the envelope:
$P^{\text{cand}}_{b,i}=\{p^{\text{model}}_{b,i}\}\cup \mathrm{Nbr}(p^{\text{model}}_{b,i})$.
The candidate set is small (typically 2--3 profiles), keeping exploration cost bounded. We treat each pair $(b,i)$ as an independent small environment and perform online learning only within $P^{\text{cand}}_{b,i}$.

The goal of online learning is not to re-fit the full latency model, but to learn residuals relative to the analytical prediction. For any candidate profile $p\in P^{\text{cand}}_{b,i}$, the analytical model predicts JCT $\hat{T}_p(c)$. Let the observed request JCT be $T^{\text{obs}}$; we define the residual as $\delta \triangleq T^{\text{obs}}-\hat{T}_p(c)$. For each candidate, we maintain an exponentially weighted moving average (EWMA) residual estimate $\bar{\delta}_{b,i}(p)$ and a usage count $N_{b,i}(p)$. After each execution, we update the residual by
\begin{equation}
\small
\label{eq:ewma_update}
\bar{\delta}_{b,i}(p)\leftarrow (1-\alpha)\bar{\delta}_{b,i}(p)+\alpha\,\delta,
\end{equation}
where $\alpha\in(0,1]$ controls tracking speed under non-stationary drift. Given $\bar{\delta}_{b,i}(p)$, the corrected effective latency is
\begin{equation}
\small
\label{eq:Teff}
T^{\text{eff}}_p = \hat{T}_p(c)+\bar{\delta}_{b,i}(p).
\end{equation}

We perform \emph{$\varepsilon$-greedy selection} over $P^{\text{cand}}_{b,i}$: with probability $1-\varepsilon$, we choose the profile that satisfies constraints and minimizes $T^{\text{eff}}_p$; with probability $\varepsilon$, we randomly explore among the remaining candidates. Because the action space per environment is at most three profiles, we do not need heavier contextual bandits (e.g., LinUCB) to achieve fast adaptation.

Online exploration carries the primary risk of SLO violations, so we enforce safety guardrails. First, we use $\hat{T}_p(c)\le T_{\text{SLO}}$ as a conservative feasibility filter; if the feasible set is empty, we fall back to a default conservative compression configuration. Second, we use a cooldown mechanism for unpredicted violations: for each profile we track recent SLO violations, and if a profile exceeds $K$ violations in the most recent $M$ uses, we temporarily remove it from the candidate set during a cooldown window to reduce repeated risk.

This online learning layer incurs negligible overhead: each request it evaluates at most 2--3 candidates and updates constant-size state, making it safe to deploy in the control plane without affecting token-level inference latency. Combined with the analytical model, the residual-corrected bandit enables KVServe to perform stable service-aware strategy selection under constraints and to sustain near-optimal end-to-end speedup under serving perturbations.

\section{Evaluation}
\label{sec:evaluation}

In this section, we structure our analysis to address the following key research questions:
\begin{itemize}[leftmargin=*]
    \item \textbf{End-to-End Performance:} How much does KVServe reduce the end-to-end completion time, compared to baselines under varying conditions? (Sec.~\ref{sec:eval_e2e})
    \item \textbf{Pareto Efficiency:} Can our offline search algorithm effectively identify the optimal compression pipelines that balance high CR with strict Acc constraints? (Sec.~\ref{sec:eval_acc_cr})
    \item \textbf{Algorithmic Effectiveness:} How do the specific optimizations in our offline search and online decision modules contribute to the overall system performance? (Sec.~\ref{sec:eval_ablation})
\end{itemize}

\subsection{Experimental Setup}
\label{sec:eval_setup}

We implement KVServe atop vLLM 0.10.1~\cite{vllm}, extending its architecture to support disaggregated prefill-decode execution with our compression pipeline injected into the communication path. Additionally, we integrate the lm-eval-harness~\cite{lm-eval} directly into the system to evaluate the accuracy impact of KV compression during online inference across \textit{PD Separation} and \textit{Prefix Caching} scenarios.

\begin{figure*}[t]
    \centering
    \includegraphics[width=\linewidth]{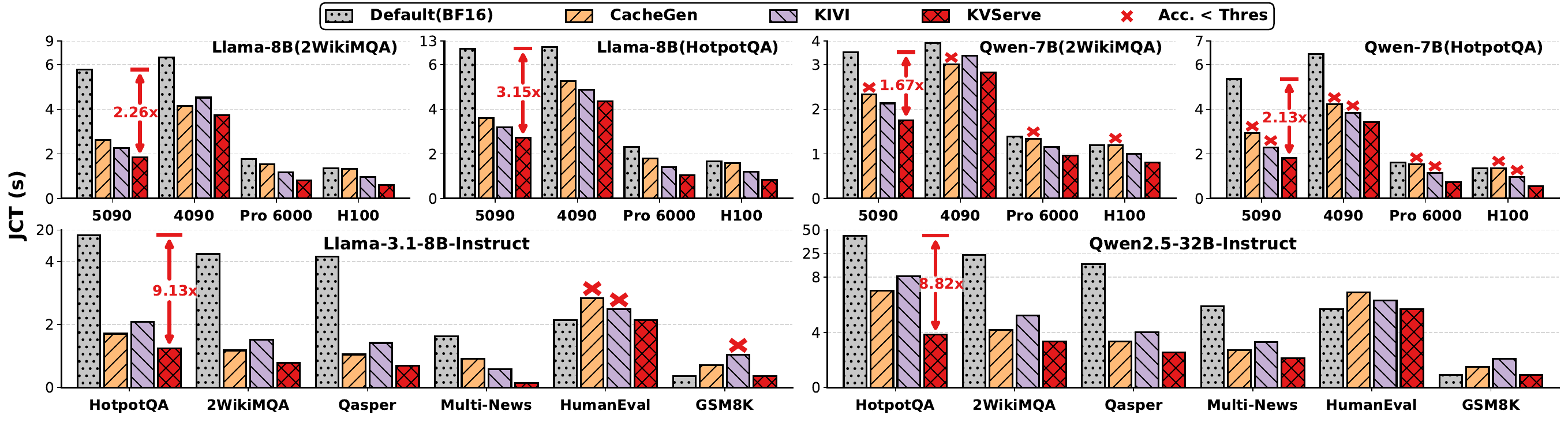} 
    \caption{End-to-End Performance across Hardware and Workloads. Top row evaluates JCT scalability across hardware tiers; bottom row benchmarks diverse datasets. Crosses (\textcolor{red}{\textbf{$\times$}}) indicate configurations failing the 97\% relative accuracy threshold.}
    \label{fig:eval_devices_datasets}
\end{figure*}

\textbf{Models and Datasets.}
We evaluate our system using Qwen2.5-7B-Instruct ~\cite{qwen2,qwen2.5}, Llama-3.1-8B-Instruct ~\cite{llama}, and the larger Qwen2.5-32B-Instruct.
Our dataset is designed to verify both search effectiveness and generalization capability: 
(i) \textit{Profiling Datasets:} We search the Pareto Frontier using four datasets: GSM8K~\cite{gsm8k} (Math), HumanEval~\cite{humaneval} (Code), Multi-News~\cite{longbench} (Summarization), and Qasper~\cite{longbench} (QA).
(ii) \textit{Unseen Datasets:} To evaluate ability, we use 2WikiMQA and HotpotQA~\cite{longbench}. These remain unseen during profiling to verify generalization to new tasks.

\textbf{Baselines.}
We compare KVServe against three optimizations, integrating core algorithms of CacheGen and KIVI as pipeline modules for comparison: 
(i) \textit{CacheGen}~\cite{Cachegen}: adapts compression by tuning quantization granularity within a fixed pipeline.
(ii) \textit{KIVI}~\cite{KIVI}: A static method applying fixed asymmetric 2-bit quantization regardless of context.
(iii) \textit{DuoAttention}~\cite{DuoAttention}: A pruning-based method benchmarking token dropping against our mixed-precision approach.

\textbf{Testbed.}
We conduct offline profiling on $4\times$ A100 (40GB) GPUs and use H100 for decoding. The prefill nodes cover three tiers with distinct network bandwidths: 
(i) \textit{Consumer Grade (10 Gbps):} $2\times$ RTX 4090 (24GB) and $2\times$ RTX 5090 (32GB).
(ii) \textit{Workstation Grade (50 Gbps):} $2\times$ RTX Pro 6000 (96GB).
(iii) \textit{Data-Center Grade (100 Gbps):} $2\times$ H100 (80GB).

\subsection{End-to-End Performance}
\label{sec:eval_e2e}

We evaluate KVServe’s JCT across diverse hardware and network configurations. Benchmarking against SOTA baselines highlights its efficiency in mitigating communication bottlenecks while preserving accuracy.

\begin{figure}[t]
    \centering
    \includegraphics[width=\linewidth]{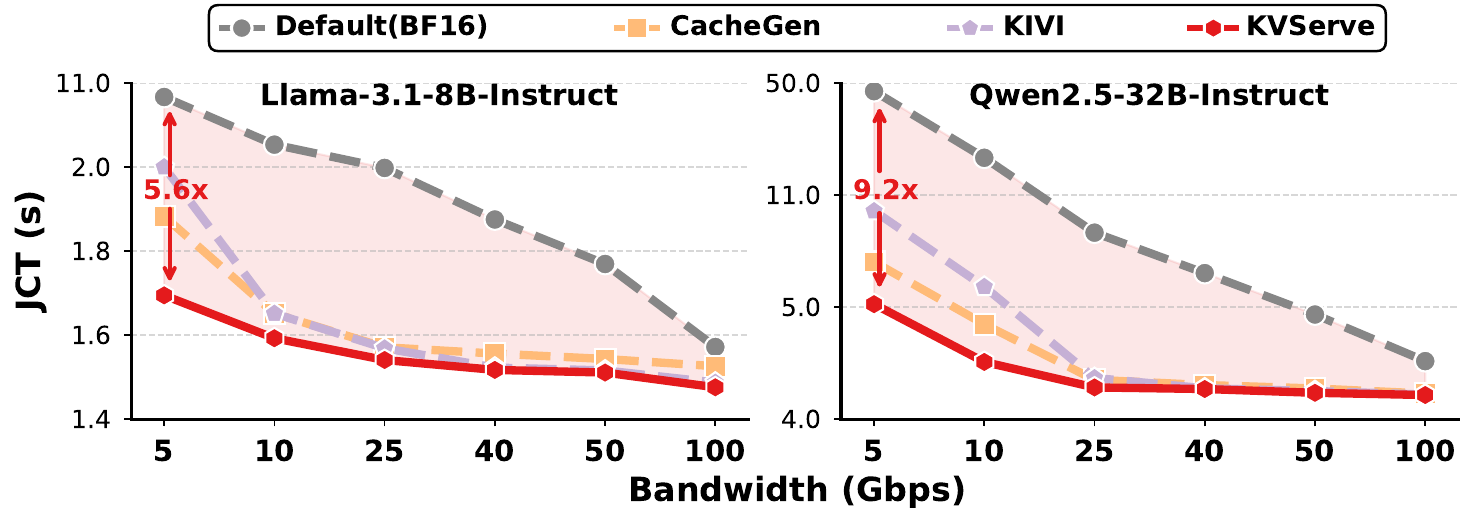} 
    \caption{JCT in PD Separation.}
    \label{fig:eval_bandwidth}
\end{figure}

\noindent\textbf{System Performance Across Diverse Hardware and Workloads.}
To assess the end-to-end performance of KVServe in practical deployment, we evaluate the JCT across a wide range of hardware tiers and diverse task categories. 

\begin{figure}[t]
    \centering
    \includegraphics[width=\linewidth]{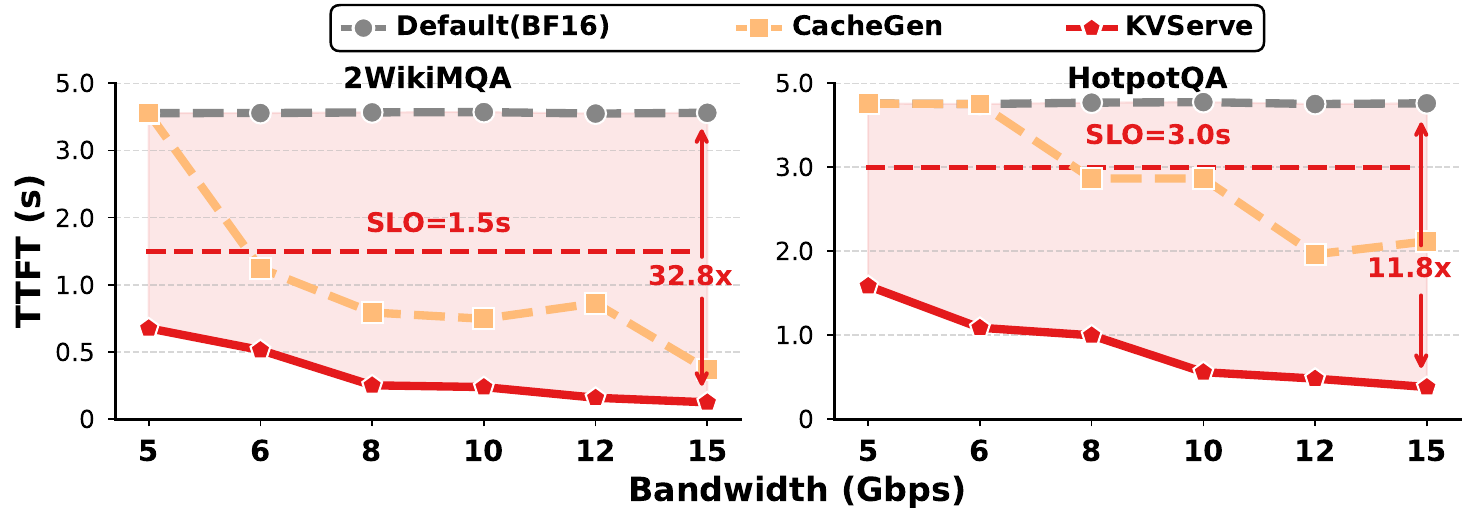} 
    \caption{TTFT in Prefix Caching.}
    \label{fig:eval_prefix_caching}
\end{figure}

As shown in the top row of Fig. \ref{fig:eval_devices_datasets}, we evaluate performance across diverse prefill hardware tiers. KVServe consistently achieves the lowest JCT, delivering up to \textit{3.15}\textbf{$\times$} speedup on bandwidth-constrained devices. Crucially, on Qwen2.5-7B-Instruct, static baselines like CacheGen and KIVI frequently violate the 97\% relative accuracy threshold (marked by \textcolor{red}{\textbf{$\times$}}), whereas KVServe strictly maintains precision while outperforming them. Even in high-bandwidth environments, KVServe avoids the significant decompression bottlenecks that plague static methods, ensuring robust performance where others often underperform.

The robustness of our system is further validated across diverse datasets using Llama and Qwen, as illustrated in the bottom row of Fig. ~\ref{fig:eval_devices_datasets}. KVServe consistently yields the lowest JCT, achieving drastic reductions on long-context tasks (e.g., \textit{9.13$\times$} on HotpotQA). A critical advantage is observed on short-context workloads like GSM8K and HumanEval, where the computational overhead of (de)compression outweighs communication savings, causing baselines to suffer negative optimization (higher JCT than Default). KVServe’s service-aware controller correctly anticipates this trade-off and bypasses compression by filtering non-beneficial profiles via theoretical modeling, ensuring performance converges to the uncompressed baseline rather than degrading it.

\begin{table*}[t]
\centering
\caption{Accuracy and Compression Efficiency. Evaluated on Qwen2.5-7B-Instruct via offline Pareto search on A100 under a 97\% relative accuracy constraint. Cell values denote Accuracy / Compression Ratio; \textbf{bold} indicates accuracy exceeding the baseline. Average Accuracy reports the relative percentage against Default.}
\label{tab:main_results_acc_cr}

\footnotesize
\setlength{\tabcolsep}{3pt} 

\begin{tabularx}{\linewidth}{p{2.5cm}|YYYY|YY|Y}
\toprule[1.5pt]
\multirow{2}{*}{\textbf{Method}} & \multicolumn{4}{c|}{\textbf{Profiling Workloads}} & \multicolumn{2}{c|}{\textbf{Unseen Workloads}} & \multirow{2}{*}{\textbf{Average}} \\
\cmidrule(lr){2-5} \cmidrule(lr){6-7}
\small{(Acc / CR)} & \textbf{GSM8K} & \textbf{HumanEval} & \textbf{Multi-News} & \textbf{Qasper} & \textbf{2WikiMQA} & \textbf{HotpotQA} & \small{(Rel. Acc / CR)} \\
\midrule[1.0pt]
\rowcolor{tbGrey} Default (BF16) & 82.64 / 1.00 & 83.54 / 1.00 & 23.73 / 1.00 & 43.34 / 1.00 & 46.96 / 1.00 & 57.53 / 1.00 & 100.00 / 1.00 \\
\midrule
CacheGen & 72.55 / 6.01 & 57.32 / 4.06 & 17.95 / 6.33 & 25.95 / 6.81 & 28.53 / 6.84 & 24.09 / 6.94 & 65.76 / 6.17 \\
KIVI & 81.50 / 4.26 & 81.71 / 2.49 & 23.38 / 4.50 & 41.05 / 4.96 & 46.33 / 5.04 & 55.37 / 5.15 & 97.43 / 4.40 \\
DuoAttention & 82.56 / 2.21 & 82.93 / 1.06 & 20.43 / 2.92 & 40.53 / 3.83 & 45.45 / 4.08 & 56.00 / 4.50 & 95.48 / 3.10 \\
\midrule
KVServe-Unified & 81.50 / 7.07 & \textbf{84.15} / \textbf{6.20} & 23.18 / 7.36 & 42.35 / 7.85 & \textbf{47.04} / \textbf{7.94} & 54.23 / 8.07 & 98.20 / 7.42 \\
\rowcolor{tbRed} \textbf{KVServe-Aware} & \textbf{84.53} / \textbf{7.29} & \textbf{84.15} / \textbf{6.04} & \textbf{24.75} / \textbf{10.12} & \textbf{43.48} / \textbf{8.60} & 46.32 / 8.72 & 55.11 / 8.90 & \textbf{100.35} / \textbf{8.28} \\
\bottomrule[1.5pt]
\end{tabularx}
\end{table*}

\noindent\textbf{Adaptive Performance Across Network Bandwidths and Serving Scenes.}
To evaluate the system adaptability under fluctuating network conditions, we analyze KVServe across two representative disaggregated scenarios: \textit{PD Separation} and state-offloading with \textit{Prefix Caching}. We enforce target bandwidths via sender-side rate control using Linux traffic shaping and NIC-level rate limiting for RoCE.

As illustrated in Fig.~\ref{fig:eval_bandwidth}, we first evaluate the end-to-end JCT in the \textit{PD Separated} serving scenario for Llama-3.1-8B-Instruct and Qwen2.5-32B-Instruct on the 2WikiMQA dataset using an Pro 6000 prefill node. Testing across bandwidths from 5 to 100 Gbps, the red shaded area highlights KVServe's substantial acceleration over the Default(BF16). Under constrained bandwidth (5 Gbps), KVServe delivers up to \textit{9.2$\times$} speedup. Notably, as bandwidth increases, KVServe maintains the optimal lower bound by dynamically selecting lower-overhead strategies, effectively avoiding the negative optimization observed in static baselines.

Beyond PD Separation setups, KVServe also excels in state-disaggregated scenarios leveraging \textit{Prefix Caching} on remote KV pools. Fig.~\ref{fig:eval_prefix_caching} depicts the Time To First Token (TTFT) for Qwen2.5-32B-Instruct on the 2WikiMQA and HotpotQA datasets using Pro 6000 node. We benchmark against CacheGen, which dynamically falls back to costly re-computation if it cannot meet the target SLO. As observed at lower bandwidths (5--6 Gbps), CacheGen fails to find a valid configuration and degrades to the Default baseline's high latency. In contrast, KVServe consistently satisfies strict SLO constraints across the entire 5--15 Gbps range by instantly pinpointing optimal profiles from its Pareto frontier. This transforms otherwise infeasible fetches into valid cache hits, achieving a peak speedup of \textit{32.8$\times$} over re-computation.

\begin{figure}[t]
    \centering
    \includegraphics[width=\linewidth]{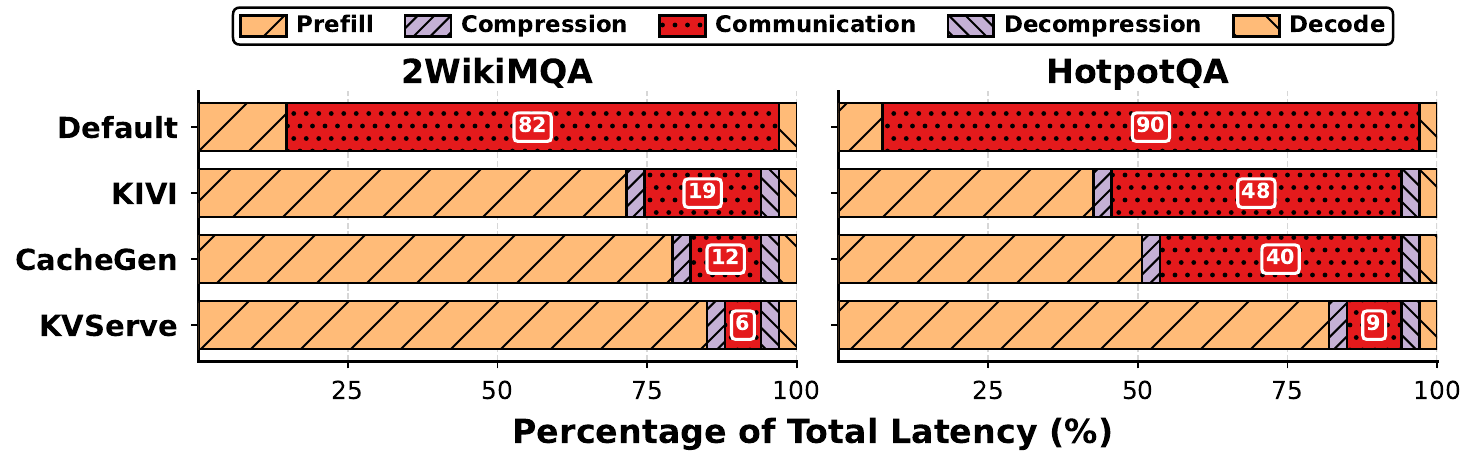} 
    \caption{Latency Breakdown across Inference Stages.}
    \label{fig:eval_breakdown}
\end{figure}

\noindent\textbf{Latency Breakdown Analysis.}
To pinpoint the source of performance gains, we decompose the end-to-end latency into five stages---Prefill, Compression, Communication, Decompression, and Decode---using Qwen2.5-32B-Instruct on 2WikiMQA and HotpotQA. As shown in Fig.~\ref{fig:eval_breakdown}, the Default baseline is severely network-bound, with communication consuming \textit{82--90\%} of the total JCT. KVServe effectively neutralizes this bottleneck, slashing the communication share to a mere \textit{6--9\%}, significantly outperforming baselines like KIVI and CacheGen in HotpotQA. The online control overhead is negligible: each decision takes $<1$\,ms. Crucially, the added computational overhead for compression and decompression remains negligible, successfully shifting the system profile from network-bound back to compute-bound.



\subsection{Accuracy and Compression Ratio}
\label{sec:eval_acc_cr}

We evaluate the quality of the compression configurations identified by our Bayesian Profiling Engine using Qwen2.5-7B-Instruct in modular pipeline. Specifically, we select the configuration that maximizes the compression ratio subject to a strict accuracy preservation constraint. Tab. ~\ref{tab:main_results_acc_cr} reports the Acc and CR across four profiling workloads and two unseen workloads. We compare two variants of our approach: \textit{KVServe-Unified}, which searches for a default robust configuration using a mixed dataset of the four profiling workloads, and \textit{KVServe-Aware}, which performs independent searches for each workload to identify the optimal configuration. For unseen workloads, \textit{KVServe-Unified} applies the configuration derived from the mixed profiling workloads, whereas \textit{KVServe-Aware} adopts the \textit{Qasper}-specific configuration due to their shared QA task alignment.

Existing methods struggle to maintain high Acc and CR on Qwen2.5-7B-Instruct. CacheGen exhibits substantial accuracy collapse across most datasets (e.g., 57.32\% on HumanEval). We attribute this to its uniform quantization; unlike Llama3, the Qwen2.5 architecture includes bias terms in Key/Value projections, resulting in a non-zero-centered, non-symmetric distribution that is ill-suited for uniform mapping. KIVI, while maintaining better stability than CacheGen, hits a compression ceiling. Although its 2-bit quantization theoretically promises an $8\times$ reduction compared to BF16, the metadata overhead required for its fine-grained group quantization limits the maximum CR to approximately $5.33\times$. Consequently, KIVI achieves an average CR of only $4.40\times$. Similarly, DuoAttention (pruning-based) fails to achieve high compression without significant loss, as aggressively discarding tokens hurts long-context retrieval accuracy.

In contrast, our profiling engine successfully navigates the trade-off space, proving robust even on unseen data. \textit{KVServe-Unified} serves as a highly effective default strategy when the workload type is unknown. By searching on a mixed dataset, it identifies a configuration that generalizes well, achieving an average CR of \textit{7.42$\times$} with a relative accuracy loss of less than \textit{2\%}. Notably, on the unseen datasets, it maintains high fidelity without any task-specific tuning, demonstrating strong ability. When the workload type is known, \textit{KVServe-Aware} unlocks superior performance by selecting specialized pipelines. It achieves an impressive average CR of \textit{8.28$\times$}---significantly outperforming all baselines---and peaks at \textit{10.12$\times$} on Multi-News. Furthermore, it maintains an average relative accuracy of \textit{100.35\%}, exceeding the Default baseline. We attribute this capability to our \textit{MixHQ} design, where the adaptive mixed-precision strategy selectively preserves significant features while filtering noise.

\begin{figure}[t]
    \centering
    \includegraphics[width=\linewidth]{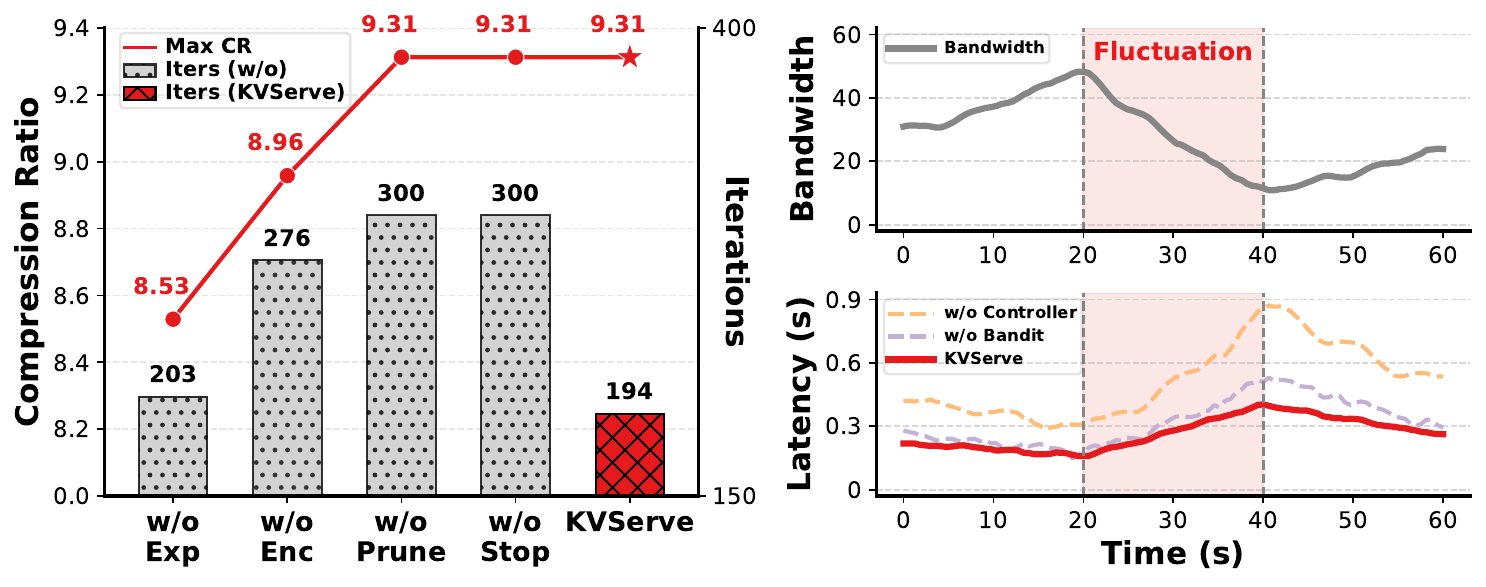} 
    \caption{Offline and Online Ablation Studies.}
    \label{fig:eval_ablation}
\end{figure}
\subsection{Ablation Studies}
\label{sec:eval_ablation}

In this section, we conduct an ablation study to decouple and quantify the individual contributions of key algorithmic components within KVServe. We specifically examine the impact of optimization strategies on the efficiency of the Bayesian Profiling Engine and evaluate the necessity of the Service-Aware Online Controller for robust performance adaptation under dynamic serving conditions.

\noindent\textbf{Efficiency of Offline Profiling Strategy.}
We evaluate the contribution of each optimization module within the Bayesian Profiling Engine by comparing the complete strategy (\textit{KVServe}) against variants excluding Heterogeneous-Parameter Encoding (\textit{w/o Enc}), Exploration-Exploitation Strategy (\textit{w/o Exp}), Bi-Directional Pruning (\textit{w/o Prune}), and Early-Stopping Mechanism (\textit{w/o Stop}). As shown in Fig.~\ref{fig:eval_ablation} (Left), removing \textit{Enc} and \textit{Exp} leads to premature convergence trapped in local optima, yielding suboptimal CR of 8.96$\times$ and 8.53$\times$, significantly lower than the global optimum of 9.31$\times$. Conversely, ablating \textit{Prune} and \textit{Stop} allows finding the optimum but fails to converge within the allocated budget, exhausting the maximum 300 iterations. The full \textit{KVServe} strategy synergizes these components, successfully identifying the global optimal configuration (\textit{9.31$\times$} CR) with superior sample efficiency, converging in just \textit{194} iterations.

\noindent\textbf{Robustness of Online Selection Policy.}
We assess the adaptability of the Service-Aware Online Controller under dynamic network conditions by monitoring end-to-end latency during bandwidth fluctuations (0--60s). The experiment compares the proposed residual-corrected approach (\textit{KVServe}) against ablations lacking the Context Bandit (\textit{w/o Bandit}) and the Online Controller (\textit{w/o Controller}). As illustrated in Fig.~\ref{fig:eval_ablation} (Right), during significant bandwidth drops (shaded area, 20s--40s), the absence of the theoretical lower-envelope model (\textit{w/o Controller}) results in severe latency spikes, peaking at nearly 0.9s, due to the selection of non-beneficial strategies. Furthermore, the lack of the online bandit (\textit{w/o Bandit}) prevents the system from correcting runtime execution drift, leading to consistently higher latency compared to the full system. In contrast, \textit{KVServe} achieves the lowest latency profile (stabilizing around \textit{0.3s}) by combining analytical modeling for baseline selection with bandit learning for real-time residual correction.

\section{Related Work}

\textbf{KV Cache Compression.} Most KV cache compression methods center on quantization. Prior work improves the accuracy--compression tradeoff by (i) reshaping KV distributions before quantization to make them more amenable to low-bit representations~\cite{QuaRot,AffineQuant,xu2025llm265,KVTC}, (ii) allocating precision at finer granularity across layers/heads/tokens/chanel to better match KV sensitivity~\cite{KIVI,Cachegen,KVQuant}, and (iii) reducing the runtime overhead of (de)compression through optimized implementations and kernels~\cite{jiang2025kvcomp,zhang2025hack}. We view these techniques as modular design knobs that can be instantiated as components and parameters in our strategy pool. In parallel, KV pruning reduces footprint by selectively retaining “important” states; it is largely orthogonal to quantization, but tends to incur larger quality loss at aggressive reduction levels~\cite{DuoAttention,KVPress,KVzap}. In contrast to KVServe, most existing approaches are \emph{service-agnostic}: they adopt fixed configurations and do not adapt to dynamic service context at runtime.

\noindent\textbf{Disaggregated Serving Optimization.}
Recent serving systems increasingly optimize \emph{disaggregated} inference. 
Phase-disaggregation systems redesign execution, scheduling, and to better utilize heterogeneous GPU pools, spanning PD separation, and scheduler-driven variants~\cite{zhong2024distserve,patel2024splitwise,feng2025windserve,sun2024llumnix,hu2025shuffleinfer,hong2025semipdefficientllmserving,duan2024muxserve}.
KV \emph{state disaggregation} and KV-pool architectures optimize KV offloading, and reuse across requests, making KV movement a first-class system concern~\cite{qin2025mooncake,chen2025impress,liu2025lmcache,li2025hotprefix}.
Elastic designs further generalize disaggregation by dynamically reallocating resources and parallelism as request mixes drift~\cite{wu2024loongserve,liu2025elasticmm,chen2025multiplexing}.
These system-level advances are complementary to our focus: we study \emph{service-aware KV compression} as an orthogonal lever that can be embedded into both PD-separated and KV-disaggregated serving stacks.

\section{Conclusion}
Disaggregated LLM serving turns the KV cache from an internal GPU state into a massive, latency-critical payload, making KV movement a dominant bottleneck. KVServe rethinks KV compression as a \emph{service-state-dependent} decision problem rather than a fixed algorithm choice. By treating KV compression as a constrained, service-dependent control problem, KVServe enables robust end-to-end speedups across both PD separation and KV state disaggregation under dynamic workloads and bandwidth. Beyond KV caching, we believe the same principle applies to a broader class of networked state-movement workloads in modern disaggregated systems—e.g., parameter offloading, and embedding retrieval. Overall, KVServe establishes a service-aware foundation for disaggregated LLM serving, showing how KV movement can be optimized as a first-class, constraint-driven control problem. \textit{This work does not raise any ethical issues.}

\begin{acks}
This work was supported by the National Natural Science Foundation of China (Grant Nos. 62032023 and T2125013), the Innovation Funding of ICT, CAS (Grant No. E461050), and the National Key Research and Development Program of China (Grant No. 2025YFB3003702). The experiments were performed on the robotic AI-Scientist platform of Chinese Academy of Sciences.
\end{acks}

\newpage
\bibliographystyle{ACM-Reference-Format}
\bibliography{refs}


\end{document}